\documentclass[11pt]{article}

\usepackage[final]{acl}

\usepackage{times}
\usepackage{latexsym}
\usepackage{amsmath}
\usepackage{amssymb}
\usepackage{booktabs} 
\usepackage{pgfplots}
\usepackage{pgfplotstable}
\usepackage{subcaption}
\usepackage{colortbl}
\usepackage{tabularx}
\usepackage{xcolor}
\usepackage{subcaption}
\pgfplotsset{compat=1.18}
\usepackage[table]{xcolor}
\usepackage{pifont}      
\usepackage{fontawesome5}

\usepackage[T1]{fontenc}

\usepackage[utf8]{inputenc}

\usepackage{microtype}

\usepackage{inconsolata}
\usepackage{multirow}

\usepackage{graphicx}

\usepackage{graphicx,transparent,pgfplots}
\pgfplotsset{compat=1.18}
\usetikzlibrary{pgfplots.groupplots}

\newcommand{\spec}[1]{\raisebox{-0.5\height}{\includegraphics[width=0.16\textwidth]{#1}}}

\title{~\raisebox{-15pt}{\includegraphics[height=40pt]{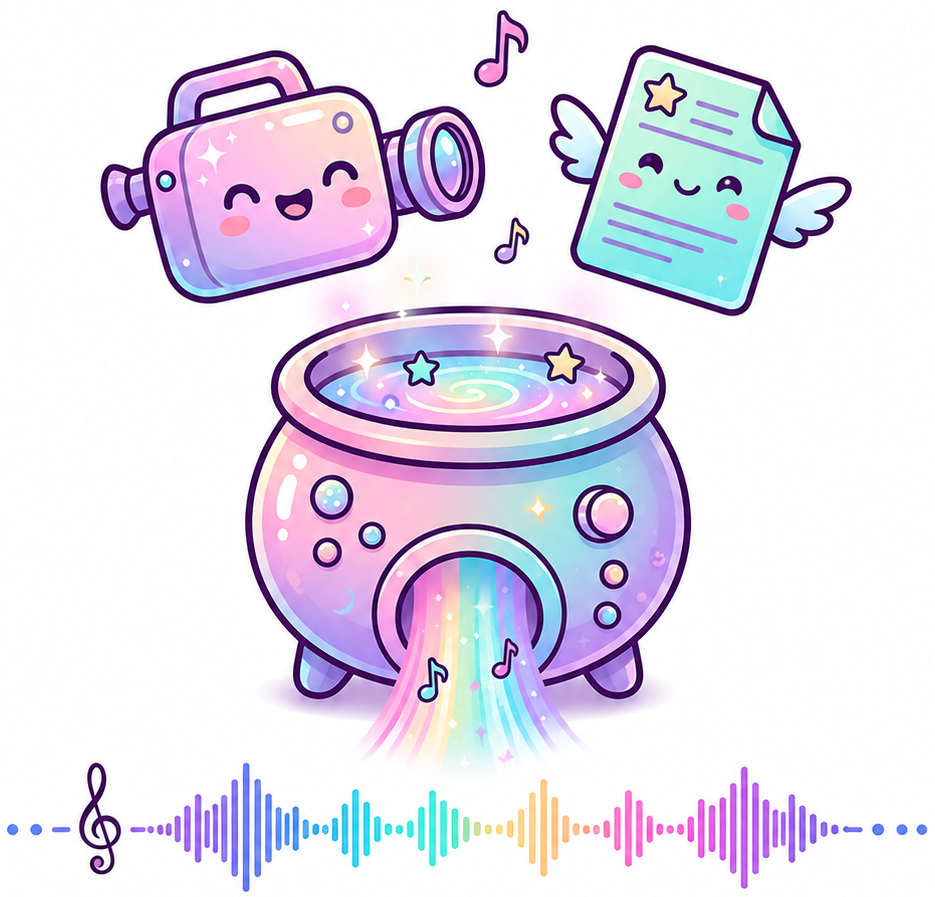}}~VIBE: Video Instruction-aligned Background music gEneration}
\author{
  Aryan Vijay Bhosale$^{1\ast}$, Vaibhavi Lokegaonkar$^{1\ast}$,
  Vishnu Raj$^{2}$, \textbf{Gouthaman KV}$^{2}$\\
  \textbf{Sreyan Ghosh}$^{1}$,
  \textbf{Ramani Duraiswami}$^{1}$,
  \textbf{Lie Lu}$^{2}$, \textbf{Dinesh Manocha}$^{1}$\\
  $^{1}${University of Maryland, College Park, USA} \quad
  $^{2}${Dolby Laboratories, USA}\\
  \texttt{\{\href{mailto:aryan.vi.bhosale@gmail.com}{aryan.vi.bhosale}, \href{mailto:vaibhavilokegaonkar@gmail.com}{vaibhavilokegaonkar}\}@gmail.com} \\ \faIcon{headphones} \href{https://vibe-text-video-to-music-generation.github.io/vibe/}{Project Page} \quad \faIcon{github} \href{https://github.com/aryanvibhosale/vibe}{Code}
}

\begin{document}
\maketitle
{\renewcommand{\thefootnote}{}\footnotetext{$^\ast$Equal contribution. }}
\begin{abstract}

Current video-to-music (V2M) models lack semantic control and fail to penalize instruction violations, largely due to their reliance on reconstruction objectives and the representational bottleneck of static cross-modal conditioning in Diffusion Autoregressive (DAR) architectures. To resolve this, we introduce VIBE, a novel text-and-video-to-music (T+V2M) generation model that leverages: (1) \textbf{Conditioning Connection}, a depth-wise cross-layer conditioning mechanism that dynamically bridges the planning and diffusion refinement heads and (2) a comprehensive \textbf{reward modeling taxonomy}, optimizing for both hard, verifiable constraints (e.g., tempo, key) and soft, subjective qualities (e.g., musicality, multimodal alignment) with a structured 5-stage training curriculum. Upon evaluation using audio-visual alignment, instruction following, and audio quality metrics, along with a subjective human evaluation study, we observe that VIBE demonstrates enhanced controllability and instruction adherence while performing comparably to most evaluated baselines on generation fidelity and multimodal alignment.
\end{abstract}

\section{Introduction}

Music and imagery jointly shape the emotional experience of short-form video, driving creators to seek background scores that align rhythmically and semantically with their visual content. Background scores, added in post-production, have been shown to reinforce the emotional tone, narrative pacing, and aesthetic of a video \cite{huh2026vidtunecreatingvideosoundtracks}. However, curating the perfect background score for a video is a highly nuanced task with no single correct answer \cite{10.1145/3715336.3735814, 10.1145/3313831.3376514}. The same video may be paired with acoustic guitar, chill lo-fi beats or suspenseful piano. Visual signals alone are insufficient to capture creator intent, and therefore, supplemental fine-grained textual prompts are required \cite{10.1145/3715336.3735814, melechovsky-etal-2024-mustango, huang2023noise2musictextconditionedmusicgeneration} to disambiguate and accurately capture creator intent.

\noindent Existing video-to-music (V2M) works such as video-only models \cite{tian2025vidmusesimplevideotomusicgeneration,zuo2025gvmgengeneralvideotomusicgeneration,ji2025diffv2mhierarchicalconditionaldiffusion}, models that accept minimal auxiliary inputs \cite{liu2024mumullamamultimodalmusicunderstanding,Kang_2024} and text$+$video-to-music models \cite{lokegaonkar2026videorobinautoregressivediffusionplanning,kim2025ossl,kim2026dialogueawarevideotomusicgenerationusing}, rely majorly on visual cues for music generation and provide limited semantic and stylistic controllability to the end user. Even when text is provided as input to Video-Robin~\cite{lokegaonkar2026videorobinautoregressivediffusionplanning}, it is used for high-level style steering only and not for fine-grained attribute specification of tempo, key, musical genre and mood. This is because most existing models are trained on reconstruction objectives over paired video-music data with no mechanism to explicitly penalise  misalignment in specific parts of the input text instruction. This lack of instruction following can be attributed to clear architectural and training limitations in such models. More specifically, this can be attributed to the architectural bottleneck of static cross-modal conditioning, especially in diffusion autoregressive models and the absence of a principled training framework for decomposed, fine-grained instruction adherence.
\vspace{0.5mm}

\noindent To overcome these issues, we present \textbf{VIBE} (\textbf{V}ideo \textbf{I}nstruction-aligned \textbf{B}ackground music g\textbf{E}neration), a novel joint text$+$video-to-music multi-preference-optimised generation technique with a structured training curriculum that provides a mechanism for dynamic conditioning to the DiT, enabling highly controllable music generation that aligns seamlessly with the textual and visual prompts. To tackle static conditioning, we propose \textit{Conditioning Connection}, a depth-wise conditioning mechanism that passes the intermediate hidden states from the Autoregressive-Head as multimodal context, along with the autoregressive audio generation history to each layer of the Refinement-Head. This provides a clear abstraction of global planning and refinement tasks across the architecture rather than collapsing the LM's representational hierarchy into a single embedding. We also address the issue that no existing video-conditioned music generation system is explicitly trained to follow individual components of music instructions, such as tempo, key, genre, and mood, as specified in the input text prompts. Our approach audits the full spectrum of musical instruction components, partitions them into objectively verifiable aspects and subjective perceptual qualities, and develops dedicated reward models for each attribute class to directly supervise instruction adherence via reinforcement learning. We also introduce a \textit{verifiability-driven taxonomy of rewards for preference optimization} combining hard signal-processing rewards for objective attributes with learned compositional multi- and omni-modal reward models for subjective attributes.
\noindent Overall, our approach can generate music that semantically 
and rhythmically adheres to fine-grained text instructions while remaining temporally coherent with the input video. Our novel contributions include:

\begin{enumerate}
    \item We propose \textbf{VIBE}, a text$+$video-to-music generation model for multimodally aligned, high fidelity music generation featuring Conditioning Connection, a novel depth-wise cross-layer conditioning mechanism that addresses the representational bottleneck of static conditioning by passing a learned linear combination of the multimodal LM's hidden states to each layer of the LocDiT, enabling rich multimodal grounding and smoother propagation of multimodal context throughout the denoising hierarchy.

    \item We introduce a systematic taxonomy of musical attributes and perform holistic reward modeling for each spanning hard, verifiable rewards for objective instructions and learned cross-modal and omni-modal reward models for subjective perceptual qualities.
    
    \item We demonstrate how the compositional reward signal can be integrated with a DiffusionNFT-style reinforcement learning objective through a comprehensive, multi-stage training recipe and precise data mixtures for preference optimization of Diffusion Autoregressive architectures under multimodal instruction-following constraints. Multiple experiments across objective metrics on Reelbench benchmarks and human evaluation demonstrate consistent gains in fine-grained instruction adherence over prior text- and video-conditioned systems, while remaining competitive on generation fidelity.
\end{enumerate}

\section{Related Work}
\subsection{Video-to-Music Generation}
Video-conditioned music generation has garnered significant research interest, with approaches centering on visual feature extraction and alignment with musical rhythm. Foundational works such as CMT \citep{Di_2021}, and Video2Music \citep{Kang_2024}, generate symbolic music in the MIDI format by using motion and semantic video features. Whereas  VidMuse \cite{tian2025vidmusesimplevideotomusicgeneration} generates high-fidelity music waveforms from video using local and global features, while MuMu-LLaMA (interchangeably referred to as M2UGen in literature) \cite{liu2024mumullamamultimodalmusicunderstanding} broadens this with visual and textual encoders to allow text prompt conditioning. Works like GVMGen \citep{zuo2025gvmgengeneralvideotomusicgeneration}, and Diff-V2M \citep{ji2025diffv2mhierarchicalconditionaldiffusion} address alignment through hierarchical attention over spatio-temporal features and specialized encoders for different aspects of music. More recently, Video-Robin \citep{lokegaonkar2026videorobinautoregressivediffusionplanning} and V2M-ZERO \citep{lin2026v2mzerozeropairtimealignedvideotomusic} have advanced the field by extending conditioning capabilities beyond discrete token spaces and eliminating the need for paired training data, respectively. A parallel line of work bridges vision and music \emph{explicitly} rather than architecturally: Visuals-Music Bridge (VMB) \citep{wang2024multimodalmusicgenerationexplicit}
converts video into textual descriptions that condition a text-to-music model, so every
visual cue must survive a discrete textual bottleneck. Conditioning Connection bridges the
two \emph{implicitly} instead, propagating a fused multimodal representation through the
depth of the diffusion stack. We test the assumption behind the VMB idea in
Section~\ref{sec:further_ablations}. Broader still, any-modality-to-audio systems such as AudioX \citep{tian2026audiox}
generate audio (including sound effects) from video, text or their combination, but target general audio
plausibility rather than adherence to specified musical attributes, placing them
adjacent to---rather than in competition with---the instruction-following setting we
study. 

\noindent Despite these advances, none of these works optimize for human perceptual preferences during training, resulting in generated music that lacks the expressivity characteristic of human-composed music. Further, capturing relative quality across a broader candidate pool and reducing dependence on large-scale preference datasets remain open challenges, as existing approaches rely on pairwise comparisons or fixed datasets. Furthermore, diffusion-based approaches such as Diff-V2M and Video-Robin suffer from static conditioning \cite{li2026semanticroutingexploringmultilayer}, where the conditioning inputs do not adapt across the depth of the Diffusion Transformer. This causes a mismatch between the requirements of each layer and the fixed conditioning signal. In this work, we address these limitations by introducing Conditioning Connection for dynamic conditioning and DiffusionNFT-based online preference optimization. 

\subsection{Preference Optimization for Conditional Music Generation}
Imbuing human preferences into the music generation process continues to be a prevalent challenge. CMI-Reward Bench \citep{ma2026cmirewardbenchevaluatingmusicreward} addresses the holistic nature of music quality evaluation by providing a reward ecosystem containing large-scale human-annotated preference datasets alongside trained reward models. Building on such foundations, recent works have explored popular preference optimisation techniques to directly align generative models with human judgement. MR-FlowDPO \citep{ziv2025mrflowdpomultirewarddirectpreference} extends DPO to flow-matching text-to-music models using multi-dimensional automated rewards across text alignment, audio quality, and semantic consistency. LeVo \citep{lei2025levohighqualitysonggeneration} constructs a semi-automatic preference dataset of approximately 60K win-lose pairs and applies interpolation-based multi-preference DPO to jointly optimise all objectives without retraining. HeartMuLa \citep{yang2026heartmulafamilyopensourced} takes a different approach, constructing separate preference datasets scored by distinct music quality metrics, training independent DPO models on each, and linearly merging the resulting checkpoints to jointly optimise style adherence, lyric clarity, and audio quality. Beyond DPO, ACE-Step v1.5 \citep{gong2025acestepstepmusicgeneration} applies GRPO to a language model planner for intrinsic reward-based alignment, deriving rewards from the model's own internal consistency, while SymphonyGen \citep{he2026symphonygen3dhierarchicalorchestral} applies GRPO with a cross-modal audio-perceptual reward to refine symbolic orchestral generation. In this work, we extend the paradigm of preference-aligned music generation conditioned on video and fine-grained text instructions.


\section{Methodology}

Generating music that simultaneously adheres to fine-grained textual instructions and aligns rhythmically with video requires solving two distinct problems. First, the architecture must propagate rich multimodal context throughout the entire denoising hierarchy. Second, the training objective must explicitly penalize violation of individual instruction attributes, a signal that reconstruction-based losses are blind to. We address both through VIBE: a Diffusion Autoregressive multimodal music generation model with Conditioning Connectors for depth-wise dynamic conditioning, trained via a structured 5-stage curriculum with diverse, composite reward formulation for preference optimization.


\subsection{Architecture}
\begin{figure}
    \centering
    \includegraphics[width=0.5\textwidth]{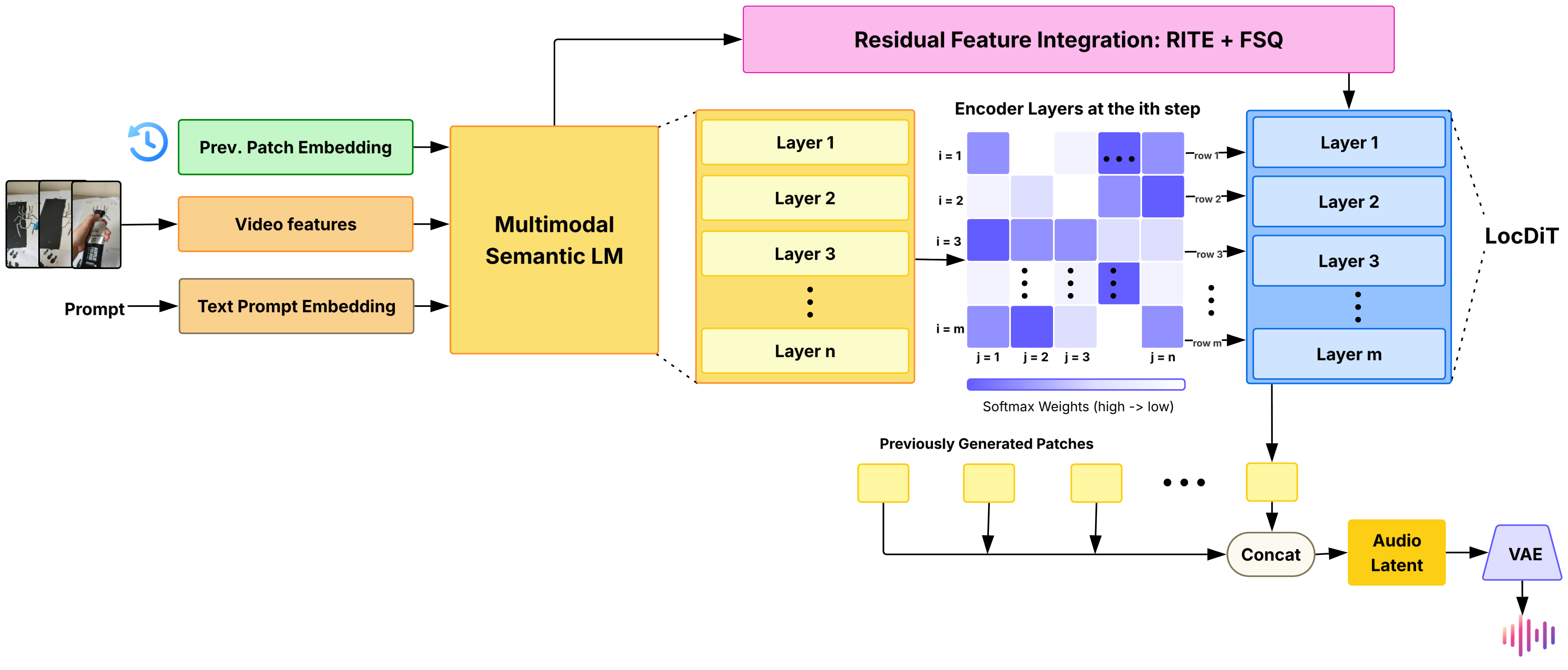}
    \caption{\textbf{VIBE Architecture.} Video frames, text prompt, and previously generated patch embeddings are passed to the Multimodal Semantic LM. Layer-wise hidden states are linearly combined via learnable per-DiT-layer coefficients to form Conditioning Connectors, which are routed to every LocDiT layer alongside residual-integrated embeddings to generate each music patch. }
    \label{fig:architecture}
\end{figure}

\noindent VIBE adopts a Diffusion Autoregressive architecture~\cite{lokegaonkar2026videorobinautoregressivediffusionplanning} that decomposes music generation into two global planning (AR-Head) and patch-wise refinement (Refinement-Head) as shown in Figure~\ref{fig:architecture}. The \textit{AR-Head} performs global semantic planning and integrates video frames (encoded by a frozen CLIP encoder), fine-grained text instructions, and the autoregressive patch history through a Multimodal Semantic LM. The resulting representation is compressed via a Finite Scalar Quantization (FSQ) bottleneck and enriched with residual acoustic detail through a Residual Integration Transformer Encoder (RITE), producing a per-patch planning embedding. The \textit{Refinement-Head}, implemented as a local Diffusion Transformer (LocDiT), denoises each latent patch conditioned on this embedding and the previously generated patch. Generated patches are re-encoded and fed back into the AR-Head autoregressively. Once all patches are complete, a pretrained VAE decoder reconstructs them as a full waveform.

\noindent In vanilla Diffusion Autoregressive models, the planning embedding is passed as a single static tensor to the LocDiT. This collapses the full representational hierarchy of the multimodal semantic LM into one embedding, ignoring the functional stratification of deep transformer stacks, wherein earlier layers encode global structural properties and later layers encode fine-grained local detail~\cite{li2026semanticroutingexploringmultilayer}. 

\noindent\textbf{Conditioning Connection.} We address this bottleneck by introducing \textbf{Conditioning 
Connection}, an architectural mechanism that carries the learned, weighted representation conditioning vectors between the multimodal semantic LM and the LocDiT. At each generation step, we compute a linear combination of hidden states from all layers of the Multimodal Semantic LM, using learnable coefficients, and pass the result to each layer of the Refinement Head. We refer to these linearly combined vectors as \textbf{Conditioning Connectors}. 

\noindent We compute a Conditioning Connector vector $\mathbf{c}^{(k)}_i$ for each LocDiT layer $k$ as a linear combination of hidden states across all LM layers using learnable coefficients:

\begin{equation}
    \mathbf{c}^{(k)}_i = \mathbf{W}\,
    \sum_{l=1}^{L} \alpha^{(k)}_l\, \mathbf{h}^{(l)}_i,
    \qquad \sum_{l=1}^{L} \alpha^{(k)}_l = 1
    \label{eq:cc}
\end{equation}

\noindent where $\alpha^{(k)}_l \in \mathbb{R}$ are learnable scalar weights specific to LocDiT layer $k$, and $\mathbf{W} \in 
\mathbb{R}^{d_{DiT} \times d_{LM}}$ is a learnable projection, $d_{DiT}$ being the hidden dimension of the LocDiT and $d_{LM}$ that for Multimodal Semantic LM. Since each autoregressive step also receives previously generated patches as input, the LocDiT layer $k$ conditions on both current multimodal context $\mathbf{c}^{(k)}_i$ \& autoregressive musical patch history $\mathbf{m}_{i-1}$ at every layer, promoting patch-to-patch continuity and strengthening rhythmic and semantic coherence across the generated sequence. 
\vspace{-5pt}
\begin{equation}
    \tilde{\mathbf{h}}^{(k)}_i = {\text{LocDiT}}^{(k)}\!\left(
        \mathbf{x}^t,\; \mathbf{c}^{(k)}_i,\; \mathbf{m}_{i-1},\; t
    \right)
    \label{eq:locdit_layer}
\end{equation}

\noindent The combination coefficients are learned independently per DiT layer, allowing each layer to attend to the level of semantic abstraction most suited to its denoising responsibility, i.e. shallower LM representations for early LocDiT layers handling global structure and deeper representations for later layers handling fine acoustic detail. Conditioning Connectors replace static single-vector cross-head conditioning with a structured representational hierarchy that flows through the full depth of the diffusion stack.

\subsection{Training Objectives}

\textbf{Pre-training and Supervised Finetuning (SFT).} Both pre-training and SFT optimize the LocDiT via a flow-matching diffusion 
loss over the velocity field $\mathbf{v}_\theta$:
\vspace{-10pt}

\begin{equation}
    \mathcal{L}_{\text{diff}} = \mathbb{E}_{t, \mathbf{x}^0, \boldsymbol{\epsilon}}
    \left\|
    \mathbf{v}_\theta(\mathbf{x}^t, \mathbf{E}_p, \mathbf{m}_{i-1}) 
    - \dot{\alpha}_t \mathbf{x}^0 - \dot{\sigma}_t \boldsymbol{\epsilon}
    \right\|_2^2
\label{eq:diff_loss}
\end{equation}

\noindent where $\mathbf{x}^t = \alpha_t \mathbf{x}^0 + \sigma_t \boldsymbol{\epsilon}$,\; $\dot{p}_t$ indicates $\frac{dp}{dt}$ for any $p$, \; $\boldsymbol{\epsilon} \sim \mathcal{N}(0, \mathbf{I})$, and $\mathbf{v}_\theta$ 
is the LocDiT velocity field. See Table~\ref{tab:notation} for notation.

\noindent\textbf{Preference Optimization.} We adopt DiffusionNFT~\cite{zheng2026diffusionnftonlinediffusionreinforcement} to perform online RL 
directly on the forward diffusion process. At each iteration, $G$ candidates 
$\{\mathbf{x}_0^g\}_{g=1}^G$ are generated per conditioning input $\mathbf{c}$ 
and scored with an optimality probability $r \in [0,1]$:

\begin{equation}
\small{
\hspace{-14pt}
    \mathcal{L}_{\text{NFT}} =
    \mathbb{E}_{\substack{\mathbf{c},\,t \\ \mathbf{x}_0 \sim \pi^{\text{old}}}}
    \!\Big[
        r \|\mathbf{v}_\theta^{+} - \mathbf{v}\|_2^2
        + (1{-}r)\|\mathbf{v}_\theta^{-} - \mathbf{v}\|_2^2
    \Big]}
    \label{eq:nft_loss}
\end{equation}

\noindent where the implicit positive and negative policies are:

\begin{align}
    \mathbf{v}_\theta^{+}(\mathbf{x}^t, \mathbf{E}_p,
\mathbf{m}_{i-1}, t) &:=
        (1{-}\beta)\,\mathbf{v}^{\text{old}} + \beta\,\mathbf{v}_\theta
        \label{eq:pos_policy} \\[2pt]
    \mathbf{v}_\theta^{-}(\mathbf{x}^t, \mathbf{E}_p,
\mathbf{m}_{i-1}, t) &:=
        (1{+}\beta)\,\mathbf{v}^{\text{old}} - \beta\,\mathbf{v}_\theta
        \label{eq:neg_policy}
\end{align}

\noindent with $\mathbf{v} = \dot{\alpha}_t \mathbf{x}^0 +
\dot{\sigma}_t \boldsymbol{\epsilon}$ the flow-matching target,
$\beta$ the guidance strength, and $\mathbf{v}^{\text{old}}$ the
frozen sampling policy. Refer to appendix for further training details.

\subsection{Reward Modelling}

We partition musical instruction attributes by verifiability into two classes: \textbf{hard verifiable} attributes (tempo and key) whose adherence is directly measurable from the audio signal, and \textbf{soft} perceptual attributes (musical genre and mood) which resist objective evaluation and require learned judgement. We develop dedicated reward models for 
each class and combine them into task-specific reward signals for preference optimization in text-to-music and video-to-music models respectively.

\noindent\textbf{Hard Verifiable Rewards.}
These rewards are developed for instruction components (tempo and key) that have discrete values and can be objectively evaluated. 

\noindent\textbf{Tempo.} We estimate the BPM $\hat{b}$ of the generated audio and parse the target tempo into one of three forms: exact value, range, or blanket descriptor (e.g.\ \textit{slow}, \textit{fast}). For exact targets we apply a Gaussian reward; for range and blanket targets a trapezoidal 
reward:
\vspace{-7.5pt}
\begin{equation}
    \boldsymbol{r}_{\text{\textbf{tempo}}} = 
    \begin{cases}
        \exp\!\left(-\dfrac{(\hat{b}-b^*)^2}{2\sigma_b^2}\right) 
        & \hspace{5pt} \text{\small \textit{exact}}\vspace{5pt}\\
        
        \max\!\left(0,\,\min\!\left(1,\,1-\dfrac{e}{\delta}\right)\right) 
        & \hspace{5pt} \text{\small \textit{range}}
    \end{cases}
    \label{eq:tempo_reward}
\end{equation}

\noindent where $e = \max(b_{\text{lo}}-\hat{b},\;\hat{b}-b_{\text{hi}},\;0)$. 
To correct for octave errors common in BPM estimators, we evaluate 
$r_{\text{tempo}}$ at $\hat{b}$, $2\hat{b}$, and $\hat{b}/2$ and take the 
maximum.

\paragraph{Key.} We average two complementary tonal alignment signals. 

\noindent The \textit{Circle-of-Fifths} (CoF) reward detects the predicted key $k_{\text{pred}}$ and penalises arc distance $d_{\text{arc}}$ from the target $k^*$, incorporating a mode penalty for major/minor mismatch:
\begin{equation}
\begin{aligned}
\Delta_k &= d_{\text{arc}}(k_{\text{pred}},k^*)
          + \mathbf{1}[\text{mode}_{\text{pred}}\neq\text{mode}^*], \\[2pt]
r_{\text{CoF}} &= s\cdot\exp\!\left(-\frac{\Delta_k^{2}}{2\sigma_k^{2}}\right)
\end{aligned}
\label{eq:cof_reward}
\end{equation}

\noindent where the reward is Gaussian weighted by the key detector confidence $s\in[0,1]$ and 
$\sigma_k=2.0$ which naturally downweights the reward when the detected key is ambiguous. 

\noindent The \textit{Krumhansl-Schmuckler} (KS) reward 
correlates the Harmonic Pitch Class Profile (HPCP) 
$\mathbf{h}\in\mathbb{R}^{12}$ against the psychoacoustically derived 
tonal hierarchy $\mathbf{p}^*\in\mathbb{R}^{12}$ for the target 
key.

\noindent The CoF reward is interpretable but depends on key detector accuracy whereas the KS reward operates directly on the spectrum and remains 
robust under detector failure.

\noindent
\begin{minipage}{0.48\linewidth}
\begin{equation}
    r_{\text{KS}} = \frac{\rho(\mathbf{h},\,\mathbf{p}^*)+1}{2}
    \label{eq:ks_reward}
\end{equation}
\end{minipage}
\hfill
\begin{minipage}{0.48\linewidth}
\begin{equation}
    \boldsymbol{r}_{\textbf{key}} = \frac{r_{\text{CoF}} + r_{\text{KS}}}{2}
    \label{eq:key_reward}
\end{equation}
\end{minipage}

\noindent\textbf{Soft Rewards.} These are developed to penalize subjective aspects of musical instruction (genre, mood) and multimodal alignment (text-music, video-music)

\noindent\textbf{Cross-Modal Reward (Text-to-Music).} We supervise musicality $r_{musicality}$ (genre, mood) and text-music alignment $r_{T\leftrightarrow M align}$ via CMI-RM~\cite{ma2026cmirewardbenchevaluatingmusicreward}, a cross-modal instruction-following reward model.

\begin{equation}
    \boldsymbol{R}^{\text{\textbf{CM}}}_{\textbf{soft}} = \frac{r_{\text{musicality}} + r_{{T\leftrightarrow M align}}}{2}
    \label{eq:soft_reward_ttm}
\end{equation}

\noindent\textbf{Omni-Modal Reward (Text$+$Video-to-Music).} Since we need to simultaneously judge musicality and alignment between text, video and generated music, we develop an omni-modal reward extracted using Qwen2.5-Omni~\cite{qwen2025qwen25technicalreport} as a judge as follows:

\begin{equation}
    \hspace{-10pt}\boldsymbol{R}^{\text{\textbf{omni}}}_{\textbf{soft}} = \frac{
        r_{\text{musicality}} + 
        r_{\text{T}\leftrightarrow\text{M} align} + 
        r_{\text{V}\leftrightarrow\text{M}align}}{3}
    \label{eq:omni_reward}
\end{equation}

\noindent\textbf{Resultant Composite Reward Signal}\\
For the text-to-music (\ref{eq:ttm_reward}) and text$+$video-to-music (\ref{eq:vtm_reward}) task respectively may be formulated as follows: 
\begin{equation}
    \boldsymbol{R}^{\text{T}\rightarrow\text{M}} = R^{\text{CM}}_{\text{soft}} + 
    \underbrace{r_{\text{tempo}} + r_{\text{key}}}_{R^{\text{hard}}}
    \label{eq:ttm_reward}
\end{equation}

\begin{equation}
    \boldsymbol{R}^{\text{T+V}\rightarrow\text{M}} = R^{\text{omni}}_{\text{soft}} + 
    \underbrace{r_{\text{tempo}} + r_{\text{key}}}_{R^{\text{hard}}}
    \label{eq:vtm_reward}
\end{equation}

\subsection{Training Curriculum}
\begin{figure}
    \centering
    \includegraphics[width=0.9\linewidth]{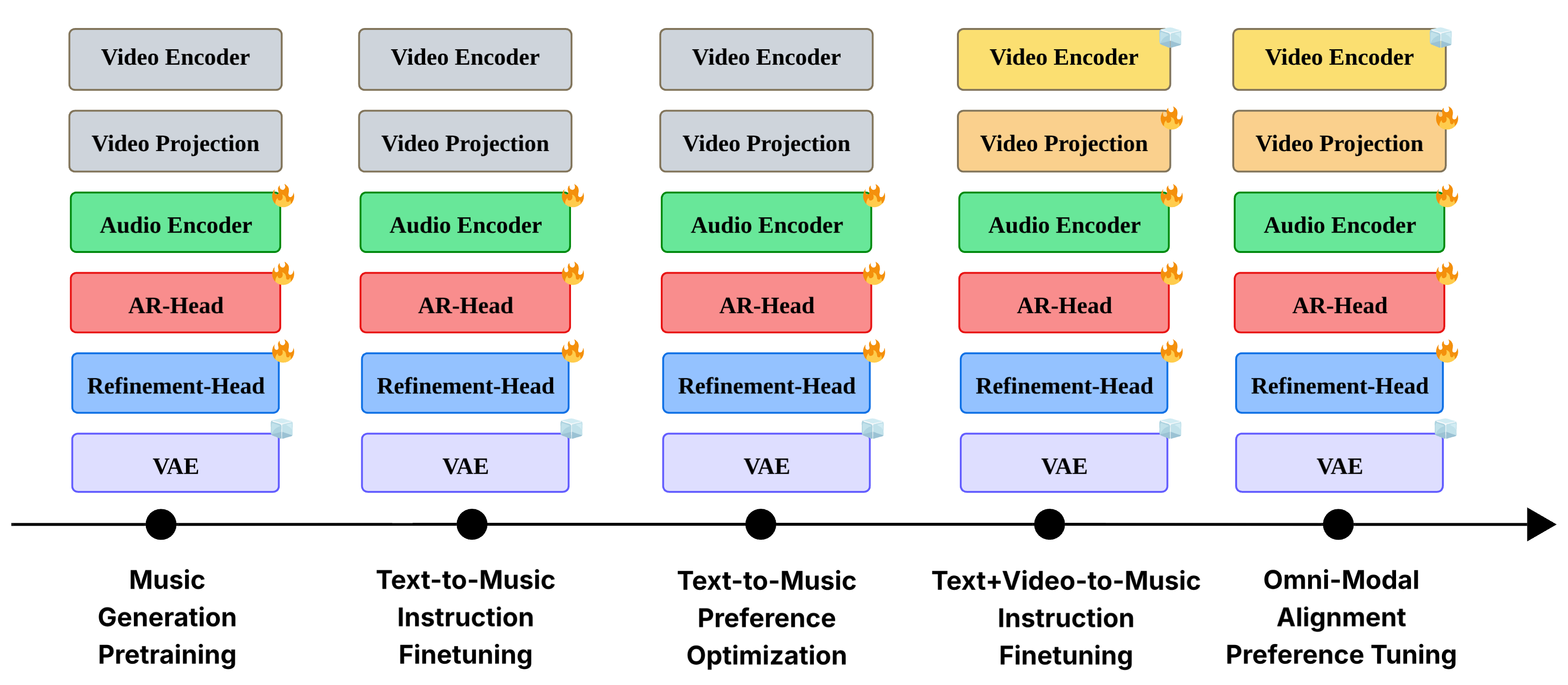}
    \caption{Overview of our training curriculum}
    \label{fig:training_curr}
\end{figure}

In this section, we describe each stage in our training curriculum. Figure \ref{fig:training_curr} illustrates this curriculum. 

\noindent\textbf{Stage 1 (Music generation pre-training)} The model is trained on large-scale text-music pairs (JamendoMaxCaps \cite{roy2025jamendomaxcaps}) without video conditioning, serving as a projector alignment phase where the AR-Head learns to map multimodal hidden states to musical latents via the Refinement-Head. This establishes a strong generative backbone for subsequent stages.


\noindent\textbf{Stage 2 (Text-to-Music Instruction Finetuning).} We finetune on a curated, instruction-rich data from MusicBench~\cite{melechovsky-etal-2024-mustango} to improve instruction-following and retain the architecture and 
training configuration as Stage 1.


\noindent\textbf{Stage 3 (Text-to-Music Preference Optimization).} We apply 
DiffusionNFT-based online RL using the composite reward $R^{\text{T}\rightarrow\text{M}}$, combining CMI-RM\cite{ma2026cmirewardbenchevaluatingmusicreward} for subjective qualities with hard verifiable rewards for tempo and key on captions from the CMI-Pref Dataset \cite{ma2026cmirewardbenchevaluatingmusicreward}. Multiple rollouts per prompt are scored and used for negative-aware finetuning, expanding the model's controllability and musicality beyond what reconstruction-based training permits.



\noindent\textbf{Stage 4 (Text$+$Video-to-Music Supervised Instruction Finetuning).} We introduce visual conditioning through the video encoder (frozen) and trainable projection layer which can enable text$+$video-to-music generation. The inputs at this stage are primarily video $+$ music pairs from HarmonySet \cite{zhou2025harmonysetcomprehensivedatasetunderstanding} and V2M \cite{tian2025vidmusesimplevideotomusicgeneration} datasets with curated fine-grained textual prompts. The projection layer is initialized and trained from scratch and while all other layers are initialized and fine-tuned from the Stage 3 checkpoint. 

\noindent\textbf{Stage 5 (Omni-modal alignment preference tuning).} 
Analogous to Stage 3, we apply DiffusionNFT using the composite V2M reward $R^{\text{T+V}\rightarrow\text{M}}$, which replaces CMI-RM with an omni-modal judge to additionally supervise video-music rhythmic and thematic alignment.

\begin{table}[h]
\centering
\resizebox{\columnwidth}{!}{%
\begin{tabular}{llr}
\toprule
\textbf{Stage} & \textbf{Training Dataset(s)} & \textbf{Size} \\
\midrule
Music Generation Pretraining     & JamendoMaxCaps   & $\approx$1.6M \\
Text-to-Music Instruction SFT & MusicBench       & 52,768        \\
Text-to-Music Preference Opt. & CMI-Pref         & 3,527         \\
Text+Video-to-Music SFT       & V2M + HarmonySet & 60,000        \\
Omni-Modal Preference Opt.    & V2M + HarmonySet & 8,000         \\
\bottomrule
\end{tabular}%
}
\caption{Training datasets used at each stage of the curriculum. All splits are training splits unless otherwise noted.}
\label{tab:training_data}
\end{table}

\section{Experiments}

\begin{table*}[t]
\centering
\resizebox{\textwidth}{!}{%
\begin{tabular}{lccccccc}
\toprule
\textbf{Model} & \textbf{FAD} $(\downarrow)$ & \textbf{FD} $(\downarrow)$ & \textbf{KL} $(\downarrow)$ & \textbf{IS} $(\uparrow)$ & \textbf{IB} $(\uparrow)$ & \textbf{Density} $(\uparrow)$ & \textbf{Coverage} $(\uparrow)$ \\
\midrule
GT
& -- & -- & -- & -- & 0.1417 & 0.9900 & 0.8800 \\
\midrule
\multicolumn{8}{c}{\textbf{Video-Only to Music}} \\
\midrule
CMT~\cite{Di_2021}
& 8.7522 & 37.7945 & 1.7329 & 1.2243 $\pm$ 0.0147 & \underline{0.1119} & 0.1084 & 0.0614 \\
GVMGen~\cite{zuo2025gvmgengeneralvideotomusicgeneration}
& 3.5729 & 16.2638 & 1.5573 & 1.7085 $\pm$ 0.0281 & 0.0957 & 0.0835 & 0.3881 \\
VidMuse~\cite{tian2025vidmusesimplevideotomusicgeneration}
& 2.3022 & 14.5385 & \underline{1.3194} & 1.4549 $\pm$ 0.0281 & \textbf{0.1233} & 0.1377 & 0.5213 \\
\midrule
\multicolumn{8}{c}{\textbf{Video + Auxiliary Inputs to Music}} \\
\midrule
Video2Music~\cite{Kang_2024}
& 22.6459 & 73.0670 & 1.8839 & 1.0233 $\pm$ 0.0014 & 0.0473 & \underline{0.1647} & 0.0084 \\
M2UGen~\cite{liu2024mumullamamultimodalmusicunderstanding}
& 4.5767 & 27.4208 & 1.5301 & 1.6499 $\pm$ 0.0567 & 0.0722 & 0.1094 & 0.2761 \\
\midrule
\multicolumn{8}{c}{\textbf{Text + Video to Music}} \\
\midrule
Video-Robin~\cite{lokegaonkar2026videorobinautoregressivediffusionplanning}
& \textbf{1.5110} & \underline{10.9020} & \textbf{1.2556} & \underline{2.0586 $\pm$ 0.0472} & 0.1017 & 0.1384 & \underline{0.5259} \\
VIBE (Ours)
& \underline{1.5829} & \textbf{10.1282} & 1.3205 & \textbf{2.3578 $\pm$ 0.0798} & 0.0888 & \textbf{0.6741} & \textbf{0.6095} \\
\bottomrule
\end{tabular}%
}
\caption{Results on the ReelBench~\cite{lokegaonkar2026videorobinautoregressivediffusionplanning} dataset. \textbf{Bold} indicates best and \underline{underline} indicates second-best.}
\label{tab:results}
\end{table*}



      

 
 
\definecolor{rowgrey}{RGB}{236, 236, 236}
\definecolor{rowblue}{RGB}{210, 222, 246}
\definecolor{rowyellow}{RGB}{252, 243, 200}
\definecolor{rowgreen}{RGB}{205, 232, 205}
\newcommand{\cmark}{\ding{51}}
\newcommand{\xmark}{\ding{55}}

\begin{table*}[t]
\aboverulesep=0pt
\belowrulesep=0pt
\renewcommand{\arraystretch}{1.25}
\centering
\small
\setlength{\tabcolsep}{4.5pt}
\resizebox{\textwidth}{!}{%
\begin{tabular}{l cccc ccccccc}
\toprule
 & \multicolumn{4}{c}{\textbf{Components}} & \multicolumn{7}{c}{\textbf{Metrics}} \\
\cmidrule(lr){2-5} \cmidrule(lr){6-12}
 & \textbf{Pre-train} & \textbf{CC} & \textbf{T2M RL} & \textbf{V2M RL}
 & \textbf{FAD}$\downarrow$ & \textbf{FD}$\downarrow$ & \textbf{KL}$\downarrow$
 & \textbf{IS}$\uparrow$ & \textbf{IB}$\uparrow$
 & \textbf{Den.}$\uparrow$ & \textbf{Cov.}$\uparrow$ \\
\midrule
\rowcolor{rowgrey}
\textit{Base architecture} & \xmark & \xmark & \xmark & \xmark
 & 3.044 & 22.592 & 1.343 & 1.588 $\pm$ 0.131 & \textbf{0.113} & 0.258 & 0.133 \\
\rowcolor{rowgrey}
\quad \textit{+ Pretraining} & \cmark & \xmark & \xmark & \xmark
 & 2.538 & 16.446 & 1.349 & 1.743 $\pm$ 0.057 & 0.0802 & 0.149 & 0.175 \\
\rowcolor{rowblue}
\quad \textit{+ Conditioning Connection} & \cmark & \cmark & \xmark & \xmark
 & 1.724 & 10.199 & 1.346 & 1.833 $\pm$ 0.069 & 0.0835 & 0.541 & 0.523 \\
\midrule
\rowcolor{rowyellow}
\textit{Preference optimization w/o CC} & \cmark & \xmark & \cmark & \cmark
 & 1.618 & 17.920 & 1.442 & 1.760 $\pm$ 0.039 & 0.0578 & 0.570 & 0.500 \\
\rowcolor{rowyellow}
\quad \textit{w/o V2M RL} & \cmark & \xmark & \cmark & \xmark
 & 1.728 & 11.372 & 1.354 & 1.763 $\pm$ 0.045 & 0.0794 & 0.290 & 0.309 \\
\rowcolor{rowgreen}
\textbf{+ CC \& preference optimization = VIBE (Ours)} & \cmark & \cmark & \cmark & \cmark
 & \textbf{1.583} & \textbf{10.128} & \textbf{1.321} & \textbf{2.358 $\pm$ 0.080} & 0.0888 & \textbf{0.674} & \textbf{0.610} \\
\bottomrule
\end{tabular}}
\caption{\textbf{Component analysis of VIBE on ReelBench.} The first block isolates the
\emph{architectural} contribution, building the model up one component at a time with no
preference optimization. The second block isolates the \emph{training} contribution, applying the
preference optimization stages (Stage 3 and Stage 5) without CC. VIBE proposes the architectural design additions and training paradigm together as a comprehensive solution for fine-grained instruction following in text+video-to-music generation. \textbf{Bold} indicates best.}
\label{tab:component_ablation}
\end{table*}

\begin{table*}[h]
\centering
\begin{tabularx}{\textwidth}{lXXXXXX|X}
\toprule
\textbf{Model} & \textbf{Rhythm} & \textbf{Theme} & \textbf{Emotion} & \textbf{Culture} & \textbf{Temporal} & \textbf{Instr. Fit} & \textbf{Overall} \\
\midrule
MuMuLLaMa & 2.490 & 2.731 & 3.000 & 3.255 & 2.143 & 3.100 & 2.724 \\
Video2Music & 2.622 & 2.744 & 2.744 & 3.293 & 2.268 & 3.090 & 2.677 \\
Video-Robin & 2.786 & 2.890 & \textbf{3.100} & 3.565 & 2.472 & 3.200 & 2.753 \\
VIBE & \textbf{2.816} & \textbf{2.910} & 3.028 & \textbf{3.581} & \textbf{2.589} & \textbf{3.240} & \textbf{2.843} \\
\bottomrule
\end{tabularx}
\caption{\textbf{Gemini omni-modal judge results.} Comparison to existing baselines along criteria defined in Table \ref{tab:gemini_axes}}
\label{tab:gemini_score_comparison}
\end{table*}

\subsection{Datasets}

We summarise the datasets used during each training stage in Table~\ref{tab:training_data}. Pretraining uses JamendoMaxCaps~\cite{roy2025jamendomaxcaps}, a large-scale instrumental music dataset. Text-to-Music Instruction Finetuning uses MusicBench~\cite{melechovsky-etal-2024-mustango}, whose tempo and key annotations support fine-grained instruction following; vocals are removed via Demucs~\cite{rouard2022hybrid}. Text-to-Music Preference Optimization uses CMI-Pref~\cite{ma2026cmirewardbenchevaluatingmusicreward}, a human-annotated preference dataset whose prompts lie in the CMI-RM training distribution. Text+Video-to-Music Instruction Finetuning samples 30,000 pairs each from V2M~\cite{tian2025vidmusesimplevideotomusicgeneration} and HarmonySet~\cite{zhou2025harmonysetcomprehensivedatasetunderstanding}; since neither provides instruction-level prompts, we use Gemini to extract tempo, key, instruments, genre, and atmosphere from each clip. Omni-Modal Alignment Preference Tuning resamples 8,000 unique videos in a 1:2 ratio of V2M to HarmonySet, providing each without its background score to Gemini-2.5-Flash to generate 4 diverse instructional prompts per video. For evaluation, we use ReelBench~\cite{lokegaonkar2026videorobinautoregressivediffusionplanning}, obtained directly from the authors.
\subsection{Implementation Details}
Our model operates in the latent space of a pretrained SongBloom~\cite{yang2025songbloom} audio VAE at 48~kHz, frozen throughout training. The SemanticLM is initialised from MiniCPM4-0.5B~\cite{minicpmteam2025minicpm4ultraefficientllmsend} with 24 transformer layers, hidden dimension 896, and 16 attention heads. Conditioning Connectors consist of 4 learnable routing vectors of size 24 and a shared 896$\rightarrow$1024 linear projection, initialised to zeros. The AR-Head uses an FSQ bottleneck with latent dimension 256 followed by an 8-layer RITE transformer. The Refinement Head is a 4-layer LocDiT~\cite{jia2025ditardiffusiontransformerautoregressive} trained with flow matching and a patch size of 4. We use frozen CLIP-ViT-Base~\citep{radford2021learningtransferablevisualmodels} as the vision encoder. For SFT stages, we use AdamW with learning rate $1\times10^{-4}$, weight decay 0.01, and warmup ratio 0.1. For preference optimization, we apply LoRA ($r{=}8$, $\alpha{=}16$) on \texttt{q\_proj} and \texttt{v\_proj} of both LM and DiT, with learning rates $2\times10^{-7}$ and $1\times10^{-7}$ respectively, a group size of 8, $\beta{=}0.5$, and 20 inference timesteps. For Stage 5, we increase LoRA rank to 64 and use Qwen2.5-Omni-7B as the multimodal judge with 4 video frames per sample.

\begin{figure}[t]
\centering
\newcommand{\mVR}[1][6mm]{\includegraphics[height=#1]{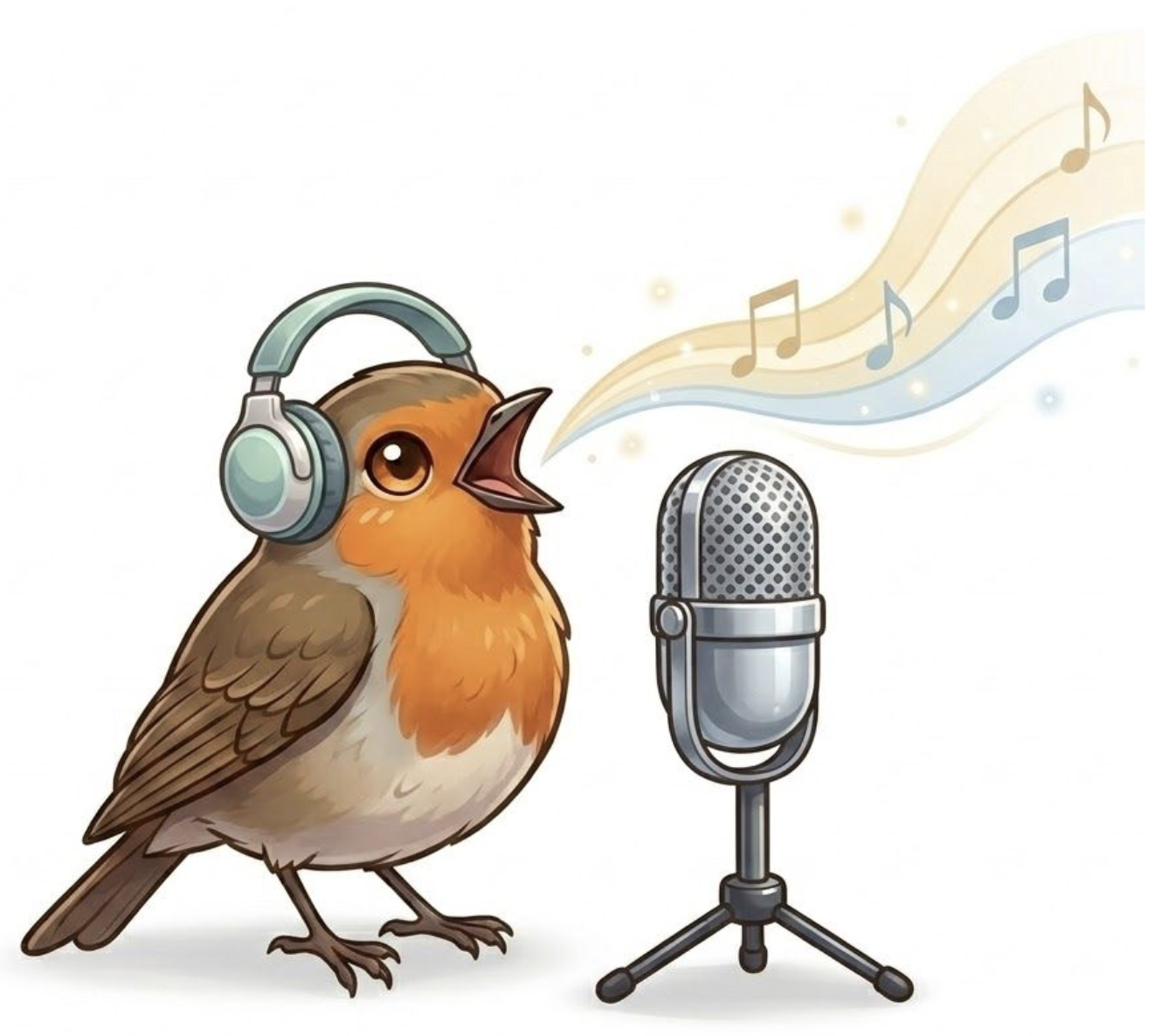}}
\newcommand{\mVB}[1][6mm]{\includegraphics[height=#1]{vibe_logo_0.png}}
\newcommand{\mAB}[1][6mm]{{\transparent{0.50}\mVB[#1]}}
\definecolor{VRor}{HTML}{B96A2B}\definecolor{VBlt}{HTML}{A98FD8}\definecolor{VBdk}{HTML}{5B3A9B}
\newcommand{\sq}[1]{\textcolor{#1}{\rule{1.4ex}{1.4ex}}}
\definecolor{VRor}{HTML}{B96A2B}\definecolor{VBlt}{HTML}{A98FD8}\definecolor{VBdk}{HTML}{5B3A9B}
\pgfplotsset{
  vibepanel/.style={
    scale only axis, width=2.85cm, height=1.95cm,
    ybar, bar width=13.5pt, bar shift=0pt, clip=true,
    xmin=-0.62, xmax=2.62, xtick={0,1,2}, xticklabels={\mVR,\mAB,\mVB},
    tick align=outside, major tick length=1.6pt,
    xtick style={draw=black!40, line width=0.4pt},
    ytick style={draw=black!40, line width=0.4pt},
    x tick label style={inner sep=0pt, yshift=-3.2pt},
    yticklabel style={font=\tiny,color=black!45,inner sep=1.6pt},
    ymajorgrids=true, grid style={line width=0.3pt, draw=black!14},
    axis x line*=bottom, axis y line*=left,
    axis line style={line width=0.4pt, color=black!25},
    title style={font=\scriptsize, yshift=-1.0ex}},
  vlab/.style={every node near coord/.append style={font=\tiny,color=black!70,anchor=south,inner sep=1.2pt}},
  vbest/.style={every node near coord/.append style={font=\tiny\bfseries,color=black,anchor=south,inner sep=1.2pt}}}
\begin{tikzpicture}
\begin{groupplot}[group style={group size=2 by 3, horizontal sep=1.0cm, vertical sep=1.5cm}, vibepanel]

\nextgroupplot[title={(a) Tempo Exact $\uparrow$}, ymin=21, ymax=25,
               ytick={21,22,23,24,25}, yticklabels={21,22,23,24,25}]
 \addplot[fill=VRor,draw=none,vlab, nodes near coords={22.0}] coordinates {(0,22.0)};
 \addplot[fill=VBlt,draw=none,vlab, nodes near coords={23.2}] coordinates {(1,23.2)};
 \addplot[fill=VBdk,draw=none,vbest,nodes near coords={24.2}] coordinates {(2,24.2)};

\nextgroupplot[title={(b) Tempo $\pm$Oct $\uparrow$}, ymin=29, ymax=32,
               ytick={29,30,31,32}, yticklabels={29,30,31,32}]
 \addplot[fill=VRor,draw=none,vlab, nodes near coords={29.7}] coordinates {(0,29.7)};
 \addplot[fill=VBlt,draw=none,vlab, nodes near coords={29.6}] coordinates {(1,29.6)};
 \addplot[fill=VBdk,draw=none,vbest,nodes near coords={31.4}] coordinates {(2,31.4)};

\nextgroupplot[title={(c) Key Exact $\uparrow$}, ymin=2.4, ymax=3.3,
               ytick={2.4,2.7,3.0,3.3}, yticklabels={2.4,2.7,3.0,3.3}]
 \addplot[fill=VRor,draw=none,vlab, nodes near coords={2.7}] coordinates {(0,2.7)};
 \addplot[fill=VBlt,draw=none,vlab, nodes near coords={2.6}] coordinates {(1,2.6)};
 \addplot[fill=VBdk,draw=none,vbest,nodes near coords={3.1}] coordinates {(2,3.1)};

\nextgroupplot[title={(d) Key Loose $\uparrow$}, ymin=9.5, ymax=11.2,
               ytick={9.5,10.0,10.5,11.0}, yticklabels={9.5,10.0,10.5,11.0}]
 \addplot[fill=VRor,draw=none,vlab, nodes near coords={10.1}] coordinates {(0,10.1)};
 \addplot[fill=VBlt,draw=none,vlab, nodes near coords={9.9}]  coordinates {(1,9.9)};
 \addplot[fill=VBdk,draw=none,vbest,nodes near coords={10.9}] coordinates {(2,10.9)};

\nextgroupplot[title={(e) Tempo MAE $\downarrow$}, ymin=32.2, ymax=33.7,
               ytick={32.2,32.7,33.2,33.7}, yticklabels={32.2,32.7,33.2,33.7}]
 \addplot[fill=VRor,draw=none,vlab, nodes near coords={33.43}] coordinates {(0,33.43)};
 \addplot[fill=VBlt,draw=none,vlab, nodes near coords={32.77}] coordinates {(1,32.77)};
 \addplot[fill=VBdk,draw=none,vbest,nodes near coords={32.42}] coordinates {(2,32.42)};

\nextgroupplot[hide axis, xmin=0, xmax=1, ymin=0, ymax=1, title={}]
 \addplot[draw=none,forget plot] coordinates {(0,0)};
 \draw[black!45, line width=0.7pt, rounded corners=2pt]
      (axis cs:0.01,0.05) rectangle (axis cs:0.99,0.95);
 \node[anchor=east] at (axis cs:0.26,0.82) {\mVR[3.5mm]};
 \node[anchor=west,font=\tiny] at (axis cs:0.31,0.82) {\sq{VRor}\, Video-Robin};
 \node[anchor=east] at (axis cs:0.26,0.50) {\mAB[3.5mm]};
 \node[anchor=west,font=\tiny] at (axis cs:0.31,0.50) {\sq{VBlt}\, VIBE w/o Stage 5};
 \node[anchor=east] at (axis cs:0.26,0.18) {\mVB[3.5mm]};
 \node[anchor=west,font=\tiny] at (axis cs:0.31,0.18) {\sq{VBdk}\, VIBE};

\end{groupplot}
\end{tikzpicture}
\caption{\textbf{Instruction-following evaluation across tempo and key metrics.} Panels (a)-(d) are higher-is-better, (e) is lower-is-better.}
\label{fig:instruction_following}
\end{figure}

\subsection{Evaluation Metrics}
We evaluate using standard audio quality metrics: \textbf{FAD}, \textbf{FD}, \textbf{KL}~\cite{zhang2023propertieskullbackleiblerdivergencemultivariate}, \textbf{IS}~\cite{salimans2016improvedtechniquestraininggans}, \textbf{Density}, and \textbf{Coverage}~\cite{ferjad2020icml}. Audio-visual alignment is measured via \textbf{ImageBind Score (IB)}~\cite{girdhar2023imagebindembeddingspacebind} and Gemini as an omni-judge \cite{liang2026omnijudgeomnillmsservehumanaligned} across \textbf{seven axes} including rhythmic sync, emotion alignment, and overall alignment. For instruction following, we report \textbf{Tempo Accuracy} and \textbf{Key Accuracy} against ReelBench ground-truth annotations (see Appendix~\ref{sec:appendix-instruct}).
\subsection{Comparison Models}

We compare against three categories of baselines. \textbf{Video-only} models include CMT~\cite{Di_2021}, GVMGen~\cite{zuo2025gvmgengeneralvideotomusicgeneration}, and VidMuse~\cite{tian2025vidmusesimplevideotomusicgeneration}. \textbf{Video + auxiliary input} models include Video2Music~\cite{Kang_2024} and M2UGen~\cite{liu2024mumullamamultimodalmusicunderstanding}, for which we extract required attributes directly from our unified text prompt. \textbf{Text + video} models include Video-Robin~\cite{lokegaonkar2026videorobinautoregressivediffusionplanning}. Models that do not support detailed text input receive the generic prompt: \textit{``Generate background music that aligns with the visual content of the video.''}

\subsection{Performance Evaluation}
In this section, we evaluate the performance of our model through ablations and comparisons to existing models detailed above. 
\subsubsection{Instruction Alignment} 
Figure~\ref{fig:instruction_following} reports instruction-following results across tempo and key metrics on ReelBench. VIBE consistently outperforms both Video-Robin and the Stage 5 ablation (VIBE w/o S5) across all five metrics, confirming that omni-modal preference tuning is the primary driver of instruction adherence. Tempo accuracy improves progressively from Video-Robin to VIBE under both exact and octave-equivalent tolerances, while Tempo MAE decreases monotonically, indicating that preference optimization produces generations that are not only more frequently correct but also closer to the target BPM on average. Key accuracy follows the same trend, with VIBE achieving the highest exact and loose key accuracy among all three models. 

\subsubsection{Quantitative comparison with baselines.}
Table~\ref{tab:results} reports results on ReelBench. VIBE achieves the best IS, FD Density, and Coverage among all models, demonstrating strong perceptual quality and generative diversity. The marginal increase in FAD and KL and increase in IB relative to Video-Robin is expected, as RL finetuning trades distributional proximity for improved perceptual quality and instruction adherence. Among baselines, VidMuse achieves competitive audio fidelity but trails on diversity, while CMT and GVMGen lag across nearly all metrics due to their lack of fine-grained conditioning. M2UGen outperforms Video2Music but remains well below VIBE, highlighting the gap between structured auxiliary inputs and free-form text conditioning paired with preference optimization.

\noindent Given the limitations of the objective metrics, we resort to 7 distinct axes to adequately measure our model’s performance following Video-Robin \cite{lokegaonkar2026videorobinautoregressivediffusionplanning}. VIBE outperforms baselines on 5 of the 6 ability axes and outperforms all evaluated baselines on overall omni-judge multimodal alignment as can be seen in Table \ref{tab:gemini_score_comparison}. The definitions of the metrics may be found in Table \ref{tab:gemini_axes}. A recent study demonstrates that omni-modal models achieve correlations to human judgments comparable to or exceeding traditional metrics (ImageBind/CLAP) on semantic alignment axes such as audio-text alignment and tri-modal coherence, citing Qwen3-Omni as an example. That being said, we would like to note that we in no way imply that such omni-judges can perfectly represent human preference and also perform subjective human evaluation as presented in Table \ref{tab:ab_test}.

\definecolor{wrbase}{RGB}{165,15,21}
\definecolor{mc1}{RGB}{0,114,178}  \colorlet{mt1}{white}
\definecolor{mc2}{RGB}{230,159,0}  \colorlet{mt2}{black}
\definecolor{mc3}{RGB}{0,158,115}  \colorlet{mt3}{white}
\definecolor{mc4}{RGB}{204,121,167}\colorlet{mt4}{black}
\definecolor{mc5}{RGB}{240,228,66} \colorlet{mt5}{black}
\definecolor{mc6}{RGB}{86,180,233} \colorlet{mt6}{black}
\definecolor{mc7}{RGB}{213,94,0}   \colorlet{mt7}{white}
\newcommand{\MID}[1]{{\setlength{\fboxsep}{1.4pt}%
  \colorbox{mc#1}{\textcolor{mt#1}{\scriptsize\bfseries#1}}}}
\newcommand{\ABtmp}{}

\newcommand{\WR}[1]{%
  \edef\ABtmp{\noexpand\cellcolor{wrbase!\the\numexpr#1*6/5-10\relax}}\ABtmp
  \ifnum#1>68\relax\color{white}\fi
  \ifnum#1>50\relax\textbf{#1}\else#1\fi}
\begin{table*}[t]
\centering
\aboverulesep=0pt \belowrulesep=0pt   
\setlength{\tabcolsep}{4pt}
\renewcommand{\arraystretch}{1.2}
\label{tab:ab_test}
\begin{subtable}[t]{0.48\textwidth}
\centering\small
\caption{Audio quality}
\begin{tabular}{l ccccccc}
\toprule
 & \MID{1} & \MID{2} & \MID{3} & \MID{4} & \MID{5} & \MID{6} & \MID{7} \\
\midrule
\MID{1}~CMT         & -- & \WR{51} & \WR{57} & \WR{38} & \WR{34} & \WR{30} & \WR{29} \\
\MID{2}~GVMGen      & \WR{49} & -- & \WR{43} & \WR{37} & \WR{42} & \WR{23} & \WR{21} \\
\MID{3}~M2UGen      & \WR{43} & \WR{57} & -- & \WR{41} & \WR{27} & \WR{34} & \WR{32} \\
\MID{4}~Video2Music & \WR{62} & \WR{63} & \WR{59} & -- & \WR{49} & \WR{30} & \WR{28} \\
\MID{5}~VidMuse     & \WR{66} & \WR{58} & \WR{73} & \WR{51} & -- & \WR{48} & \WR{46} \\
\MID{6}~Video-Robin & \WR{70} & \WR{77} & \WR{66} & \WR{70} & \WR{52} & -- & \WR{30} \\
\MID{7}~VIBE (ours) & \WR{71} & \WR{79} & \WR{68} & \WR{72} & \WR{54} & \WR{70} & -- \\
\bottomrule
\end{tabular}
\end{subtable}
\hfill
\begin{subtable}[t]{0.48\textwidth}
\centering\small
\caption{Musicality}
\begin{tabular}{l ccccccc}
\toprule
 & \MID{1} & \MID{2} & \MID{3} & \MID{4} & \MID{5} & \MID{6} & \MID{7} \\
\midrule
\MID{1}~CMT         & -- & \WR{47} & \WR{59} & \WR{40} & \WR{40} & \WR{25} & \WR{24} \\
\MID{2}~GVMGen      & \WR{53} & -- & \WR{39} & \WR{37} & \WR{38} & \WR{33} & \WR{31} \\
\MID{3}~M2UGen      & \WR{41} & \WR{61} & -- & \WR{41} & \WR{29} & \WR{38} & \WR{38} \\
\MID{4}~Video2Music & \WR{60} & \WR{63} & \WR{59} & -- & \WR{37} & \WR{34} & \WR{32} \\
\MID{5}~VidMuse     & \WR{60} & \WR{62} & \WR{71} & \WR{63} & -- & \WR{43} & \WR{41} \\
\MID{6}~Video-Robin & \WR{75} & \WR{67} & \WR{62} & \WR{66} & \WR{57} & -- & \WR{47} \\
\MID{7}~VIBE (ours) & \WR{76} & \WR{69} & \WR{62} & \WR{68} & \WR{59} & \WR{53} & -- \\
\bottomrule
\end{tabular}
\end{subtable}
\par\vspace{16pt}
\begin{subtable}[t]{0.48\textwidth}
\centering\small
\caption{Video-music alignment}
\begin{tabular}{l ccccccc}
\toprule
 & \MID{1} & \MID{2} & \MID{3} & \MID{4} & \MID{5} & \MID{6} & \MID{7} \\
\midrule
\MID{1}~CMT         & -- & \WR{51} & \WR{45} & \WR{57} & \WR{28} & \WR{17} & \WR{16} \\
\MID{2}~GVMGen      & \WR{49} & -- & \WR{43} & \WR{53} & \WR{36} & \WR{25} & \WR{23} \\
\MID{3}~M2UGen      & \WR{55} & \WR{57} & -- & \WR{55} & \WR{27} & \WR{31} & \WR{30} \\
\MID{4}~Video2Music & \WR{43} & \WR{47} & \WR{45} & -- & \WR{29} & \WR{27} & \WR{25} \\
\MID{5}~VidMuse     & \WR{72} & \WR{64} & \WR{73} & \WR{71} & -- & \WR{41} & \WR{39} \\
\MID{6}~Video-Robin & \WR{83} & \WR{75} & \WR{69} & \WR{73} & \WR{59} & -- & \WR{31} \\
\MID{7}~VIBE (ours) & \WR{84} & \WR{77} & \WR{70} & \WR{75} & \WR{61} & \WR{69} & -- \\
\bottomrule
\end{tabular}
\end{subtable}
\hfill
\begin{subtable}[t]{0.48\textwidth}
\centering\small
\caption{Overall}
\begin{tabular}{l ccccccc}
\toprule
 & \MID{1} & \MID{2} & \MID{3} & \MID{4} & \MID{5} & \MID{6} & \MID{7} \\
\midrule
\MID{1}~CMT         & -- & \WR{53} & \WR{53} & \WR{50} & \WR{34} & \WR{19} & \WR{18} \\
\MID{2}~GVMGen      & \WR{47} & -- & \WR{43} & \WR{44} & \WR{34} & \WR{30} & \WR{28} \\
\MID{3}~M2UGen      & \WR{47} & \WR{57} & -- & \WR{45} & \WR{27} & \WR{31} & \WR{30} \\
\MID{4}~Video2Music & \WR{50} & \WR{56} & \WR{55} & -- & \WR{29} & \WR{30} & \WR{28} \\
\MID{5}~VidMuse     & \WR{66} & \WR{66} & \WR{73} & \WR{71} & -- & \WR{41} & \WR{39} \\
\MID{6}~Video-Robin & \WR{81} & \WR{70} & \WR{69} & \WR{70} & \WR{59} & -- & \WR{35} \\
\MID{7}~VIBE (ours) & \WR{82} & \WR{72} & \WR{70} & \WR{72} & \WR{61} & \WR{65} & -- \\
\bottomrule
\end{tabular}
\end{subtable}
\caption{\textbf{Results of the Human A/B evaluation.} Win rates (\%) from pairwise subjective comparisons across 4 criteria. Fleiss $\kappa$ = 0.249, Krippendorff $\alpha$ = 0.245. Each cell is the win rate of the row model
against the column model. Darker shade indicates higher win rate.} 
\label{tab:ab_test}
\end{table*}

\subsubsection{Human Evaluation: A/B Testing.}
We conducted A/B testing (shown in Table~\ref{tab:ab_test}) with 18 participants aged 18--30, representative of the short-form video audience. Each evaluator compared music from two randomly selected models paired with the same video across four criteria, yielding four binary judgements per comparison. Across 20 videos and 7 models, each pair was assessed by three judges with majority vote determining the win. The criteria, also adopted in \cite{tian2025vidmusesimplevideotomusicgeneration} are: \textbf{Audio Quality} (signal fidelity and absence of artifacts), \textbf{Musicality} (standalone coherence of melody, harmony, and rhythm), \textbf{Video-Music Alignment} (temporal and semantic correspondence between music and video), and \textbf{Overall Assessment} (holistic audiovisual preference).

\subsection{Ablations}

\paragraph{Effect of removing Conditioning Connections.} Table~\ref{tab:component_ablation} demonstrates the importance of the conditioning connection. Adding it to the pretrained model reduces FAD by 32.07\%. The base architecture attains a marginally higher IB, but its poor FAD indicates the model does not possess strong priors for high fidelity music generation.
\paragraph{Stage-based ablations.} Table \ref{tab:component_ablation} shows the effects of removing certain training stages from the pipeline. The table shows that all the training stages combined with conditioning connection provide us with better performance on most audio quality metrics. Further, Table~\ref{tab:ace_step_comparison} shows our Stage~3 model, a text-to-music generation model, (text-to-music Preference Optimization) outperforms ACE-Step~v1.5~\cite{gong2025acestepstepmusicgeneration} when given video captions as input instructions. 

\paragraph{Visuals Music Bridge (VMB).}
\noindent To evaluate our omni-modal reward model for text+video-to-music generation, we compare it against a cross-modal baseline inspired by Visual-Music-Bridge (VMB) \cite{wang2024multimodalmusicgenerationexplicit} due to the lack of dedicated reward models in this space. VMB generates detailed video-music alignment captions using Gemini-2.5-Flash and fine-tunes a text-to-music model on them, assuming the text fully captures the visual context. As shown in Table \ref{tab:visual_bridge_ablation}, our model trained using Qwen2.5-Omni judge clearly outperforms VMB across all qualitative metrics.

\section{Conclusion}
We presented VIBE, a text-and-video-to-music model that addresses static cross-modal conditioning and the absence of instruction-following supervision via Conditioning Connection and a structured 5-stage training curriculum with holistic reward modeling. Experiments demonstrate strong performance on perceptual quality, generative diversity, and instruction-following metrics, with human evaluation confirming preference for VIBE across all assessment axes.

\section*{Limitations}
VIBE inherits the frozen VAE and encoder constraints of its base architecture, which may limit expressivity in niche genres. ImageBind, used as an audio-visual alignment proxy, is not trained on music data and may not fully capture semantic correspondence between generated music and video. VIBE currently targets 10-second instrumental clips and does not support vocal music, long-form scoring, or interactive editing. Finally, genre and mood rewards rely entirely on CMI-RM, whose own limitations propagate into the reward signal, and omni-modal reward extraction via Qwen2.5-Omni adds non-trivial inference overhead during Stage 5 training.

\section*{Acknowledgement}
The research at the University of Maryland is partially supported by Adobe, Amazon, NVIDIA, and Sesame.

\bibliography{custom}

\appendix

\section*{Appendix}
\label{sec:appendix}

\section{Potential Risks}
Potential Risks
VIBE generates instrumental background music conditioned on video and text, which introduces several potential risks that warrant consideration.
Misuse for Unauthorized Content. Automated music generation could be used to produce background scores for videos containing harmful, misleading, or manipulative content. Music is a powerful emotional amplifier, and pairing generated scores with propaganda, misinformation, or deceptive advertising could enhance the persuasive impact of such material.
Copyright and Intellectual Property. Although VIBE generates novel audio rather than retrieving existing tracks, models trained on large-scale music corpora may inadvertently reproduce melodic fragments, harmonic progressions, or timbral characteristics closely resembling copyrighted works. Users may unknowingly deploy generated music that infringes on existing intellectual property, particularly in commercial contexts.
Economic Displacement. Scalable, high-quality background music generation may reduce demand for human composers and music licensors in the short-form video ecosystem. While the tool is intended to complement creative workflows, widespread adoption without appropriate safeguards could disproportionately affect independent musicians who rely on licensing income from stock music and background scoring.
Bias in Reward Models. Our reward modeling framework relies on CMI-RM and Qwen2.5-Omni as judges for subjective musical qualities. These models may encode cultural, stylistic, or genre biases present in their training data, potentially favoring Western tonal music conventions and underrepresenting non-Western musical traditions. This could lead VIBE to systematically produce outputs that lack diversity in cultural representation.
Deepfake and Deceptive Media. Combined with advances in video generation and voice synthesis, automated music scoring could lower the barrier to producing convincing synthetic audiovisual media. We encourage the development of provenance tracking and watermarking mechanisms for AI-generated music to mitigate this risk.

\section{License of Artifacts}
JamendoMaxCaps is derived from the Jamendo platform, where tracks are released under Creative Commons licenses permitting research use. MusicBench is released under a CC-BY 4.0 license. CMI-Pref and CMI-RM are released for research purposes by their authors. V2M and HarmonySet are publicly available research datasets; we use them in accordance with their stated terms. ReelBench was obtained directly from its authors with permission for benchmarking. Pre-trained model weights are used under their respective licenses: MiniCPM4-0.5B under the Apache License 2.0, CLIP under the MIT License, SongBloom under its research license, and Qwen2.5-Omni under the Qwen License Agreement. The source code released is extended from VoxCPM under it's Apache License. All artifacts are used in compliance with their distribution terms for non-commercial academic research.

\section{Use of AI Assistants}
We used AI assistants at several stages of this work, detailed below.
Data Preparation. In Stage 4 (Text+Video-to-Music Instruction Finetuning), we used Google Gemini to extract structured musical attributes — including tempo, key, instruments, genre, and atmosphere — from audio clips in the V2M and HarmonySet datasets, which do not provide instruction-level text prompts natively. In Stage 5 (Omni-Modal Alignment Preference Tuning), we used Gemini-2.5-Flash to generate 4 diverse instructional prompts per video for the 8,000 resampled training videos. All generated captions were used as training inputs and were not presented as human annotations.
Reward Extraction. During preference optimization, we used Qwen2.5-Omni-7B as an omni-modal judge to extract reward signals for musicality, text-music alignment, and video-music alignment (Section 3.3). CMI-RM was used as a cross-modal reward model for text-to-music quality assessment. Both models were used programmatically as scoring functions within the training loop, not for subjective editorial decisions.
Visuals Music Bridge (VMB) Baseline. For the VMB ablation (Section 4.6), we used Gemini-2.5-Flash to generate diverse music captions containing video-music alignment details for each training video.
Writing Assistance. We used AI language models (ChatGPT and Claude) for proofreading, grammar correction, and minor stylistic editing of the manuscript text. All scientific content, experimental design, analysis, and claims were produced entirely by the authors. No AI assistant was used to generate or fabricate experimental results.

\section*{Training Objective Details}

\paragraph{Notation.} Table~\ref{tab:notation} defines all symbols used 
in Methodology.

\paragraph{Forward process.} The noisy latent at timestep $t$ follows 
$\mathbf{x}^t = \alpha_t \mathbf{x}^0 + \sigma_t \boldsymbol{\epsilon}$, 
with $\boldsymbol{\epsilon} \sim \mathcal{N}(0, \mathbf{I})$. The 
flow-matching target velocity is the time derivative 
$\mathbf{v} = \dot{\alpha}_t \mathbf{x}^0 + \dot{\sigma}_t\boldsymbol{\epsilon}$.

\paragraph{Optimality probability mapping.} Given group rewards 
$\{R^g\}_{g=1}^G$, we compute normalised advantages 
$\hat{R}^g = (R^g - \mu)/(\sigma + \varepsilon)$, clip to $[-A, A]$, 
and map to $r^g = \hat{R}^g/(2A) + 0.5 \in [0,1]$.

\paragraph{KL regularisation.} The penalty 
$\|\mathbf{v}_\theta - \mathbf{v}^{\text{ref}}\|_2^2$ approximates 
a KL divergence in velocity space and 
prevents reward hacking without requiring full distribution estimation.

\begin{table}[h]
\centering
\small
\begin{tabular}{ll}
\toprule
\textbf{Symbol} & \textbf{Description} \\
\midrule
$\mathbf{x}^0$ & Clean latent patch \\
$\mathbf{x}^t$ & Noisy latent at timestep $t$ \\
$\mathbf{v}_\theta$ & LocDiT velocity field (training policy) \\
$\mathbf{v}^{\text{old}}$ & Frozen sampling policy (EMA-updated) \\
$\mathbf{v}^{\text{ref}}$ & Frozen reference policy (for KL) \\
$\mathbf{E}_p$ & Multimodal planning embedding \\
$\mathbf{m}_{i-1}$ & Previously generated patch \\
$\alpha_t, \sigma_t$ & Flow schedule signal/noise coefficients \\
$\beta$ & NFT guidance strength \\
$r$ & Per-sample optimality probability \\
$A$ & Advantage clip threshold \\
$\lambda_{\text{KL}}$ & KL regularisation weight \\
\bottomrule
\end{tabular}
\caption{Notation used in Methodology}
\label{tab:notation}
\end{table}

\section{Instruction-Following Evaluation Protocol}
In figure \ref{fig:instruction_following}, Tempo is evaluated at exact ($\pm$10\%) and octave-equivalent tolerances alongside mean absolute error in BPM; key is evaluated at exact and loose tolerances, where the loose criterion additionally accepts relative and parallel key matches.
\label{sec:appendix-instruct}
\begin{table}[h]
\centering
\small
\begin{tabular}{ll}
\toprule
\textbf{Note Name} & \textbf{Pitch Class} \\
\midrule
C              & 0  \\
C\#\ /\ D$\flat$  & 1  \\
D              & 2  \\
D\#\ /\ E$\flat$  & 3  \\
E              & 4  \\
F              & 5  \\
F\#\ /\ G$\flat$  & 6  \\
G              & 7  \\
G\#\ /\ A$\flat$  & 8  \\
A              & 9  \\
A\#\ /\ B$\flat$  & 10 \\
B              & 11 \\
\bottomrule
\end{tabular}
\caption{Pitch class mapping used for key estimation. Enharmonic equivalents share the same pitch class index.}
\label{tab:pitch_class}
\end{table}

\begin{table}[h]
\centering
\small
\resizebox{\columnwidth}{!}{
\begin{tabular}{lcccccccccccc}
\toprule
\textbf{Mode} & \textbf{C} & \textbf{C\#} & \textbf{D} & \textbf{D\#} & \textbf{E} & \textbf{F} & \textbf{F\#} & \textbf{G} & \textbf{G\#} & \textbf{A} & \textbf{A\#} & \textbf{B} \\
\midrule
Major & 1 & 0 & 1 & 0 & 1 & 1 & 0 & 1 & 0 & 1 & 0 & 1 \\
Minor & 1 & 0 & 1 & 1 & 0 & 1 & 0 & 1 & 1 & 0 & 1 & 0 \\
\bottomrule
\end{tabular}
}
\caption{Chromagram templates for major and minor modes used in key estimation. Each binary vector encodes the scale degrees present in the corresponding mode, rooted at C. For estimation in other keys, the template is circularly shifted by the target pitch class.}
\label{tab:chroma_templates}
\end{table}

\subsection*{Tempo Evaluation}
We estimate the tempo of each generated clip in beats per minute (BPM) using \texttt{librosa}'s beat tracker. Given a ground-truth tempo $b^*$ and a predicted tempo $\hat{b}$, we evaluate correctness under two criteria. \textbf{Exact Tempo Accuracy} considers a prediction correct if $|\hat{b} - b^*| / b^* \leq 0.10$, i.e., within a 10\% relative tolerance. \textbf{Octave-Equivalent Tempo Accuracy} additionally accepts predictions at half or double the ground-truth tempo, accounting for common octave errors in BPM estimators; a prediction is accepted if any of $\hat{b}$, $2\hat{b}$, or $\hat{b}/2$ falls within the 10\% tolerance of $b^*$. We also report \textbf{Tempo MAE}, the mean absolute error in BPM across all samples with valid ground-truth tempo annotations.

\begin{table*}
\centering
\scriptsize
\setlength{\tabcolsep}{3.5pt}
\renewcommand{\arraystretch}{1.0}

\resizebox{\textwidth}{!}{%
\begin{tabular}{lccccccc}
\toprule
& \multicolumn{7}{c}{Metrics} \\
\cmidrule(lr){2-8}
Ablation & FAD$\downarrow$ & FD$\downarrow$ & KL$\downarrow$ & IS$\uparrow$ &
IB$\uparrow$ & Den.$\uparrow$ & Cov.$\uparrow$ \\
\midrule
VIBE w/o Soft Rewards   & 1.6231 & 10.1979 & 1.3394 & 1.6607 $\pm$ 0.0388 & 0.0792 & 0.1431 & 0.2191 \\
VIBE w/o Hard Rewards   & 1.6738 & 10.2975 & 1.3617 & 1.6317 $\pm$ 0.0352 & 0.0796 & 0.152 & 0.2132 \\
VIBE & \textbf{1.5829} & \textbf{10.1282} & \textbf{1.3205} & \textbf{2.3578 $\pm$ 0.0798} & \textbf{0.0888} & \textbf{0.6741} & \textbf{0.6095} \\
\bottomrule
\end{tabular}%
}
\caption{Ablation study on reward design in \textbf{VIBE}. We evaluate the contribution of soft and hard reward components to generation quality and alignment performance.}
\label{tab:reward_ablations}
\end{table*}

\begin{table*}[t]
\centering
\resizebox{\textwidth}{!}{%
\setlength{\tabcolsep}{9pt}
\renewcommand{\arraystretch}{1.0}

\begin{tabular}{lccccccc}
\toprule
& \multicolumn{7}{c}{Metrics} \\
\cmidrule(lr){2-8}
Model & FAD$\downarrow$ & FD$\downarrow$ & KL$\downarrow$ & IS$\uparrow$ & IB$\uparrow$ & Den.$\uparrow$ & Cov.$\uparrow$ \\
\midrule
Visual Music Bridge   & 2.2713 & 18.8573 & 1.4181 & 1.5457 $\pm$ 0.0998 & 0.0163 & 0.0 & 0.0825 \\
VIBE (Ours) & \textbf{1.5829} & \textbf{10.1282} & \textbf{1.3205} & \textbf{2.3578 $\pm$ 0.0798} & \textbf{0.0888} & \textbf{0.6741} & \textbf{0.6095} \\
\bottomrule
\end{tabular}%
}
\caption{Comparison between \textbf{VIBE} and the Visual Music Bridge Technique. We evaluate the impact of the visual bridging strategy on generation quality and video-music alignment metrics.}
\label{tab:visual_bridge_ablation}
\end{table*}

\definecolor{cellgreen}{RGB}{212, 237, 218}
\definecolor{cellred}{RGB}{248, 215, 218}

\begin{table*}[t]
\centering
\small
\setlength{\tabcolsep}{9pt}
\renewcommand{\arraystretch}{1.1}
\begin{tabular}{c p{0.30\textwidth} c c c c}
\toprule
\textbf{Spectrogram} & \textbf{Instruction (abridged)} &
\textbf{Target} & \textbf{Measured} &
\textbf{Target} & \textbf{Measured} \\
 & & \textbf{Tempo} & \textbf{Tempo} & \textbf{Key} & \textbf{Key} \\
\midrule
\spec{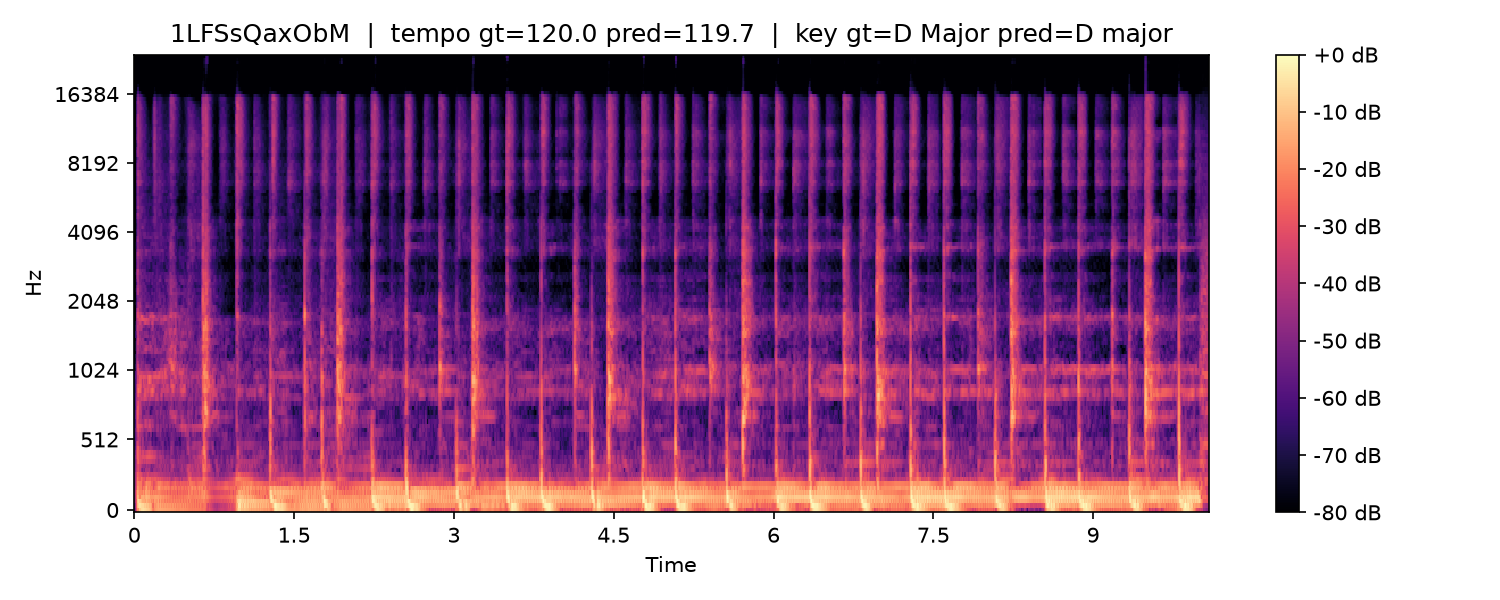} & 120 BPM Chillwave in D Major; lush evolving pads, soft kicks, crisp hi-hats
  & \cellcolor{cellgreen}120.0 & \cellcolor{cellgreen}119.7
  & \cellcolor{cellgreen}D Major & \cellcolor{cellgreen}D Major \\
\midrule
\spec{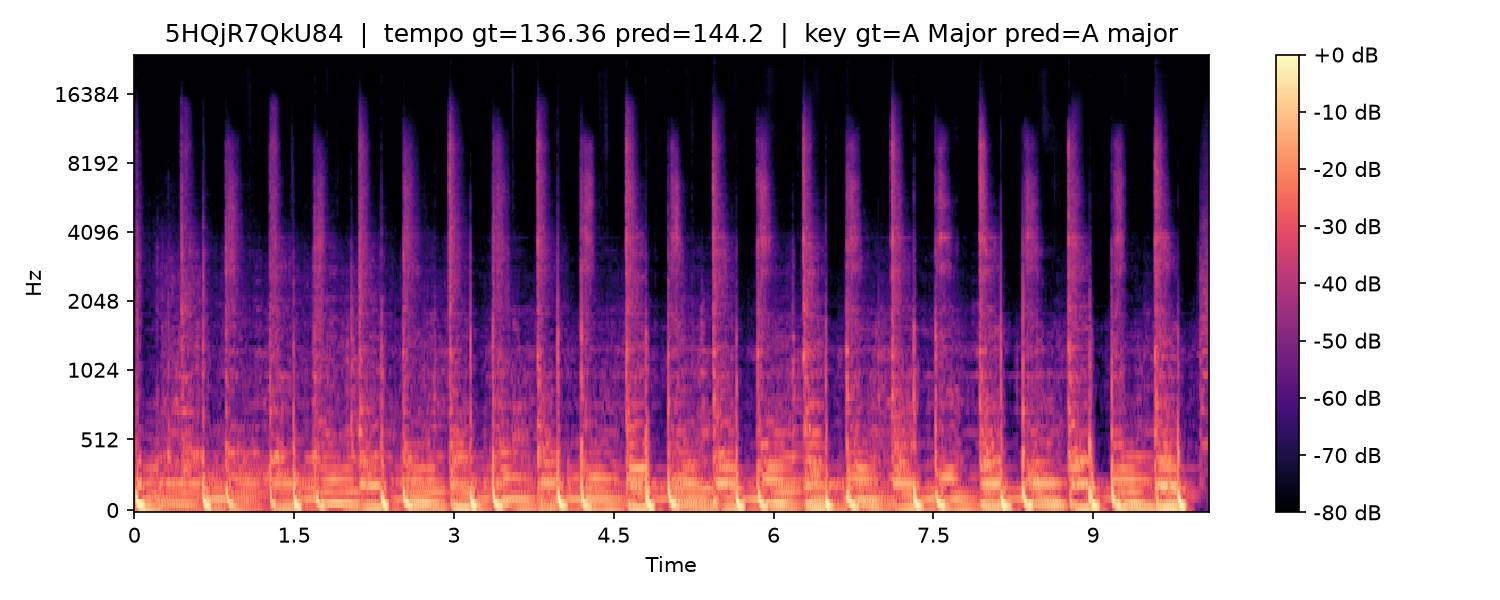} & Ambient-electronic, 136.36 BPM in A Major; evolving pads, delicate arpeggios
  & \cellcolor{cellgreen}136.4 & \cellcolor{cellgreen}144.2
  & \cellcolor{cellgreen}A Major & \cellcolor{cellgreen}A Major \\
\midrule
\spec{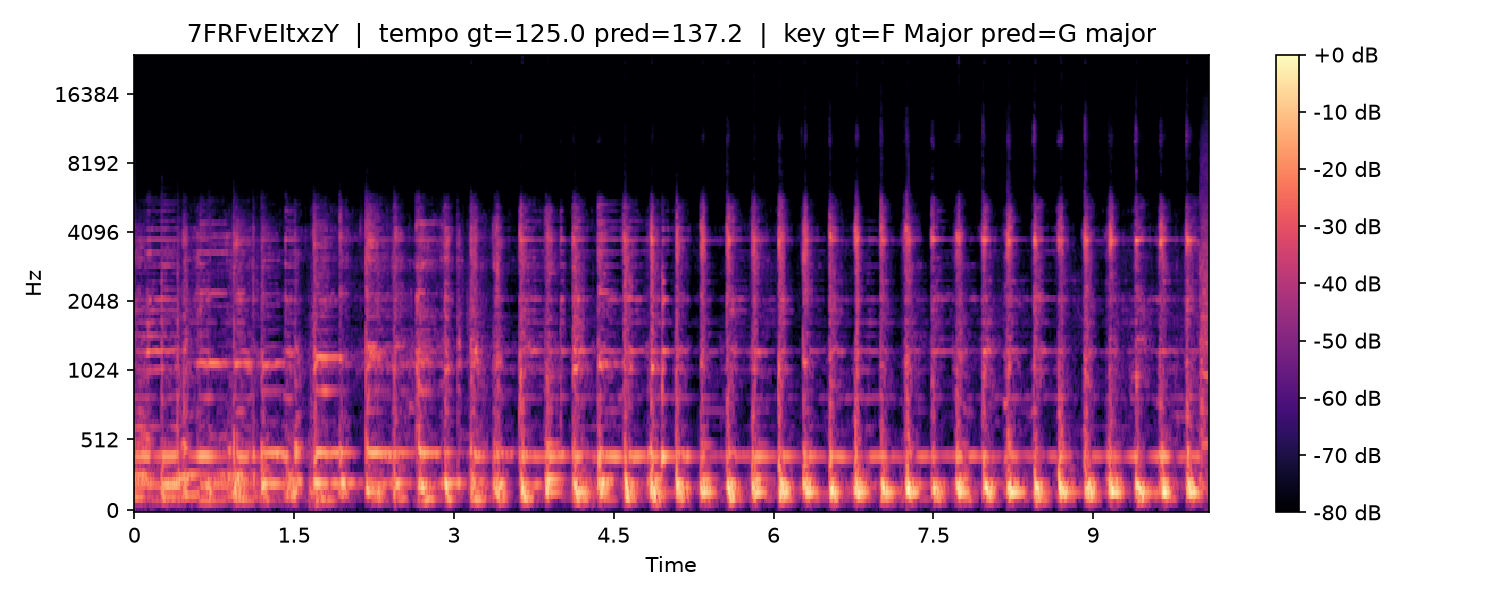} & Progressive Trance, 125 BPM in F Major; four-on-the-floor kick, side-chained bass, arpeggiated leads
  & \cellcolor{cellgreen}125.0 & \cellcolor{cellgreen}125.0
  & \cellcolor{cellred}F Major & \cellcolor{cellred}E Major \\
\midrule
\spec{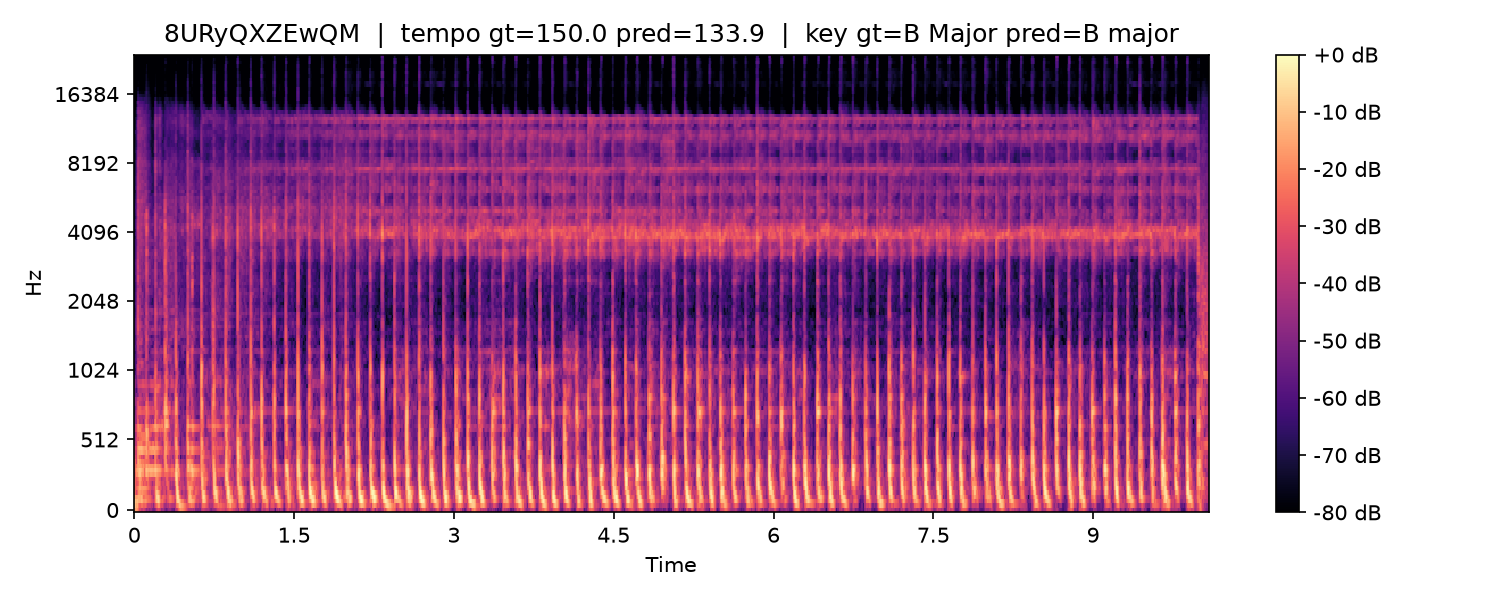} & Ethereal ambient-electronic, 150 BPM in B Major; evolving pads, arpeggiated electric guitar
  & \cellcolor{cellred}150.0 & \cellcolor{cellred}133.9
  & \cellcolor{cellgreen}B Major & \cellcolor{cellgreen}B Major \\
\midrule
\spec{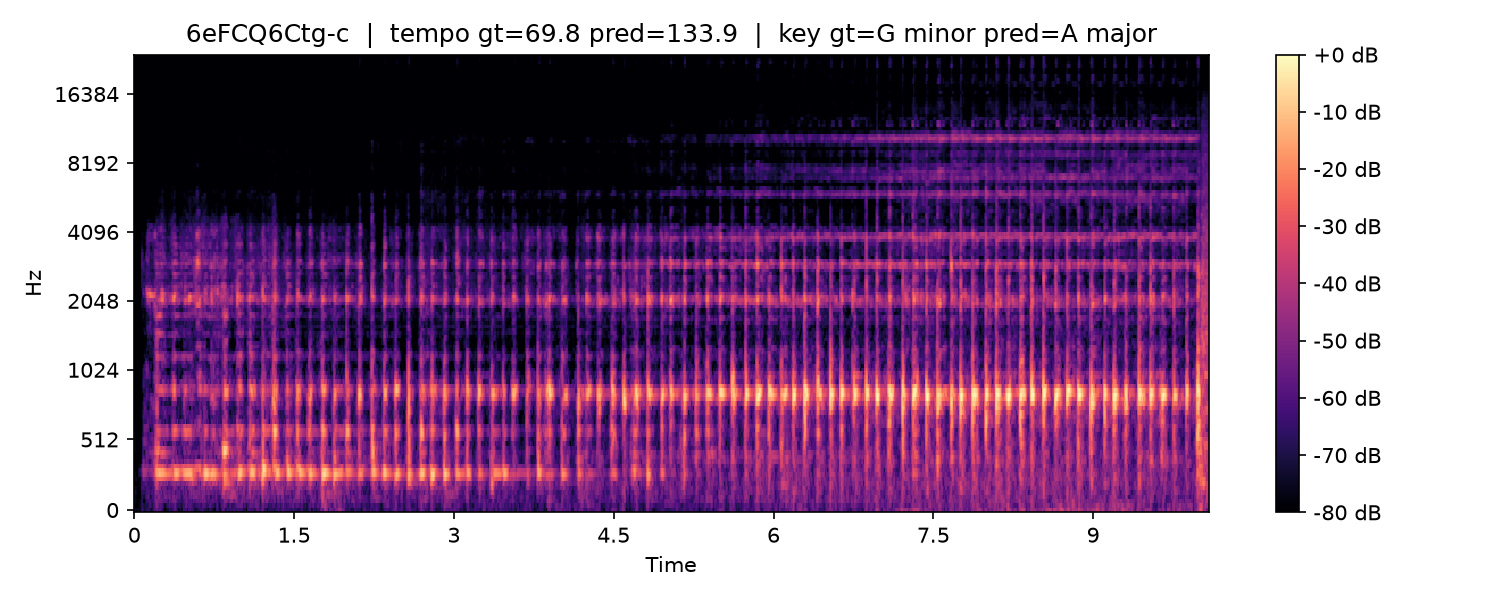} & Slow atmospheric ambient, $\approx$69.8 BPM in G minor; evolving pads, sustained bass
  & \cellcolor{cellred}69.8 & \cellcolor{cellred}133.9
  & \cellcolor{cellred}G Minor & \cellcolor{cellred}A Major \\
\bottomrule
\end{tabular}
\caption{Qualitative case studies showing different failure modes of fine-grained instruction adherence on HarmonySet test videos, with mel spectrograms of the generated clips. Green cells: the generated clip matches the instructed attribute (tempo within $\pm$10\% of target; key requiring identical tonic and mode). Red cells: a miss. The two successful cases span the tolerance band (0.3\% and 5.8\% tempo error). The last row's tempo matches only under octave equivalence and is marked as a miss under the exact criterion. Tempo (BPM) and key measured with the estimators of Appendix D.}
\label{tab:case_studies}
\end{table*}

\subsection*{Key Evaluation}
We estimate the key of each generated clip by computing a chromagram using a constant-Q transform via \texttt{librosa}, averaging chroma energy across frames to obtain a 12-dimensional pitch class profile, and correlating it against major and minor key templates across all 12 pitch classes. The pitch class mapping used for parsing ground-truth key annotations is shown in Table~\ref{tab:pitch_class}, where enharmonic equivalents such as C\#\ and D$\flat$ are treated as identical. The chromagram templates for major and minor modes are shown in Table~\ref{tab:chroma_templates}; for a key rooted at pitch class $p$, the template is circularly shifted by $p$ positions before computing cosine similarity against the predicted chroma profile. The key with the highest similarity is taken as the predicted key.

Given a predicted key $(k_{\text{pred}}, m_{\text{pred}})$ and ground-truth key $(k^*, m^*)$, where $k$ denotes the pitch class and $m \in \{\text{major}, \text{minor}\}$ denotes the mode, we evaluate under two criteria:

\begin{itemize}
    \item \textbf{Exact Key Accuracy}: the prediction is correct if $k_{\text{pred}} = k^*$ and $m_{\text{pred}} = m^*$.
    \item \textbf{Loose Key Accuracy}: the prediction is accepted under any of three conditions: (i) exact match, (ii) relative key match, where a predicted major key accepts a ground-truth minor key a minor third above ($(k^* - k_{\text{pred}}) \bmod 12 = 9$) or vice versa ($(k^* - k_{\text{pred}}) \bmod 12 = 3$), or (iii) parallel key match, where $k_{\text{pred}} = k^*$ but $m_{\text{pred}} \neq m^*$.
\end{itemize}

Both metrics are computed only over samples for which a valid ground-truth key annotation is available in ReelBench.

\section{Further Ablations and associated details}
\label{sec:further_ablations}
\subsection{Importance of Hard and Soft Rewards}
In each of the RL stages, we use the Hard Verifiable and Soft rewards for penalizing the specific musical attributes not followed from the music prompt into the rendered audio. Table \ref{tab:reward_ablations} show that better audio quality is observed when both the kinds of rewards are used.  

\subsection{Visuals Music Bridge \cite{wang2024multimodalmusicgenerationexplicit}}
\begin{equation}
    R^{\text{T+V$\rightarrow$M}}_{VMB} = R^{\text{CM}}_{\text{soft}} + 
    \underbrace{r_{\text{tempo}} + r_{\text{key}}}_{R^{\text{hard}}}
\end{equation}

\noindent Each training video is captioned offline by Gemini-2.5-Flash into four diverse music prompts; at each RL step one caption is sampled per candidate and scored by CMI-RM, providing a text-grounded proxy for video-music alignment without invoking the video encoder during 
training.

\section{Qualitative Comparisons and Failure Modes}
Qualitative examples for music generation can be see at our \href{https://github.com/VIBE-text-video-to-music-generation/vibe}{project page}. Further, we analyze instruction following ability of VIBE and discuss failure modes in Table~\ref{tab:case_studies}

\section{Human Study Instructions}
\begin{figure*}
    \centering
    \includegraphics[width=0.8\linewidth]{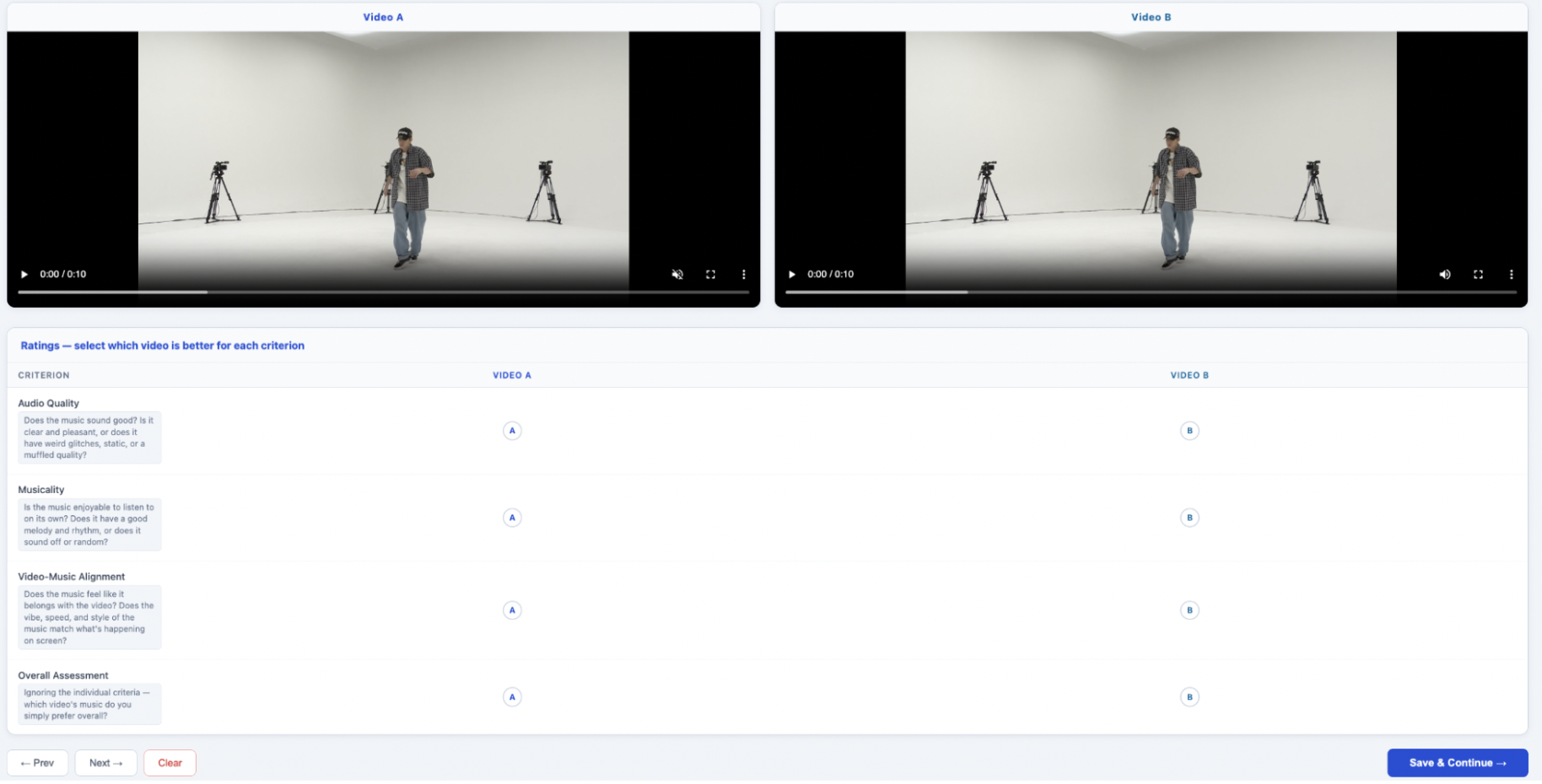}
    \caption{Sample of the study shown to  participants for A/B Testing results as shown in Table \ref{tab:ab_test}}
    \label{fig:sample_pair}
\end{figure*}

Figure  \ref{fig:sample_pair} depicts the sample shown to the participants for review. Participants were unpaid volunteers recruited from within our institution aged 18-30.

\begin{table*}
\centering
\resizebox{\textwidth}{!}{%
\begin{tabular}{lrrrrrrr}
\toprule
Model & FAD $\downarrow$ & FD $\downarrow$ & KL $\downarrow$ & IS $\uparrow$ & IB $\uparrow$ & Density $\uparrow$ & Coverage $\uparrow$ \\
\midrule
ACE-Step1.5 & 8.642 & 31.727 & 1.667 & 1.799 $\pm$ 0.056 & 0.083 & 0.140 & 0.061 \\
VIBE (Stage 3) (Ours) & \textbf{3.963} & \textbf{24.465} & \textbf{1.503} & \textbf{1.900 $\pm$ 0.079} & \textbf{0.097} & \textbf{0.325} & \textbf{0.094} \\
\bottomrule
\end{tabular}%
}

\caption{\textbf{Ablation proving the importance of video conditioning.} Comparing text-to-music models on the video-to-music generation task  when given video captions as input. This is an essential experiment to understand whether visual signals from the video are truly necessary for multimodal music generation or the context provided by the video can be substituted in text form just as effectively. Our text-to-music (TTM) model outperforms current SOTA TTM model ACE-Step1.5.}
\label{tab:ace_step_comparison}
\end{table*}

\begin{figure*}[t]
    \centering
    \includegraphics[width=0.75\linewidth]{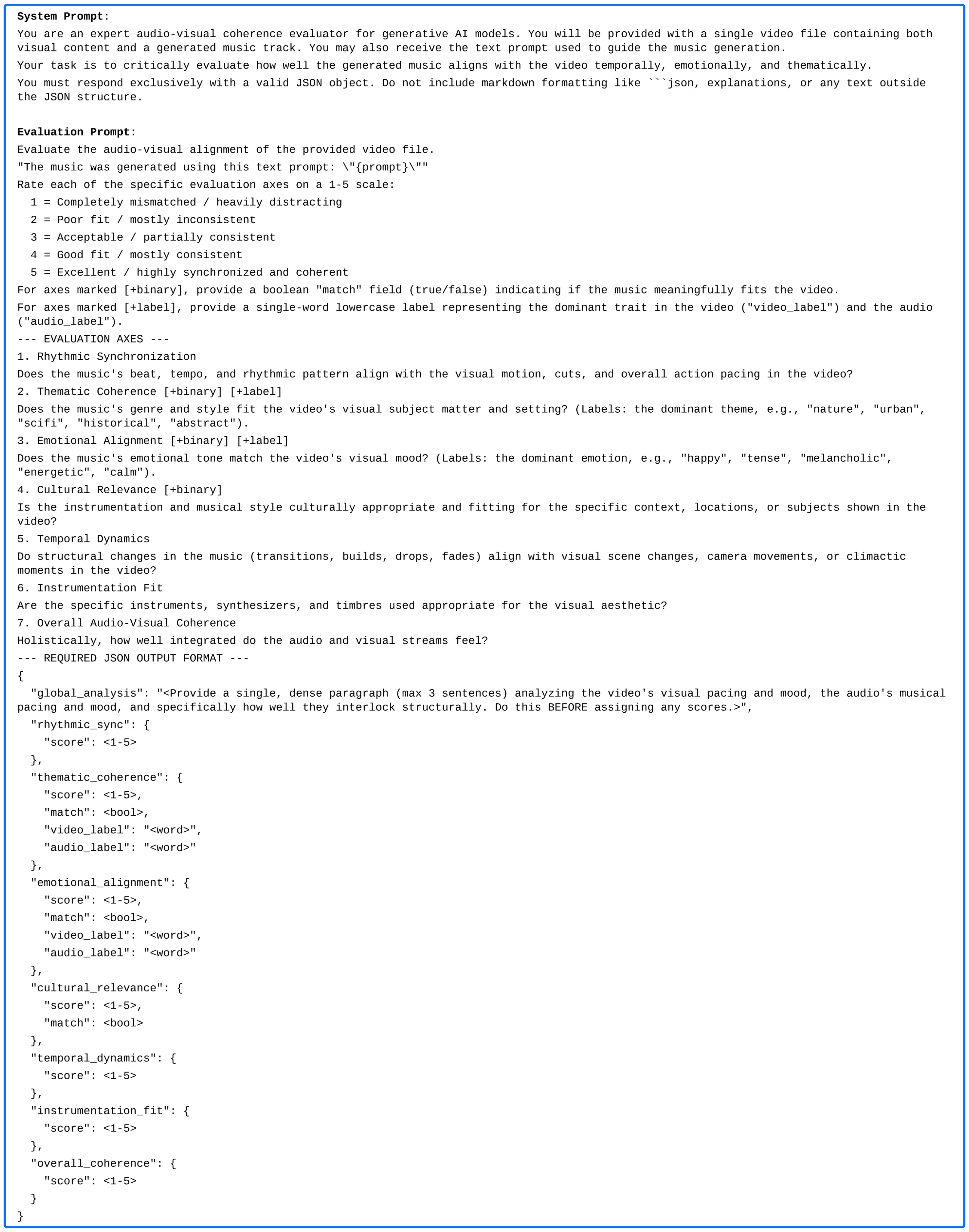}
    \caption{System prompt and evaluation prompt used to configure Gemini as an Omni-Judge for audio-visual alignment evaluation \cite{lokegaonkar2026videorobinautoregressivediffusionplanning}.}
    \label{fig:gemini_prompt}
\end{figure*}

\begin{table*}[t]
\centering
\small
\begin{tabular}{l p{0.78\textwidth}}
\toprule
\textbf{Axis} & \textbf{Definition} \\
\midrule
Rhythmic Sync & Whether beat structure, tempo and rhythmic patterns align with the motion dynamics, visual cuts and pacing of the video, i.e.\ low-level temporal synchronisation between musical rhythm and visual events. \\
Theme Coherence & Whether the genre and stylistic character of the music suit the subject matter and setting. The judge also predicts a dominant single-word theme per modality (e.g.\ \textit{nature}, \textit{urban}, \textit{sci-fi}) and reports whether the two agree. \\
Emotion Alignment & Whether the emotional tone of the music matches the visual mood. The judge predicts a dominant emotion per modality (e.g.\ \textit{happy}, \textit{tense}, \textit{melancholic}, \textit{calm}) and reports whether they correspond. \\
Cultural Relevance & Whether instrumentation, musical style and sonic motifs are culturally appropriate to the setting shown; culturally specific scenes may call for regionally relevant styles or instruments. \\
Temporal Dynamics & Whether structural changes in the music---builds, drops, transitions, fades---coincide with salient visual events such as scene changes, camera movement or climactic moments. \\
Instrumentation Fit & Whether the particular instruments, synthesisers and timbres suit the visual aesthetic and narrative context of the video. \\
Overall Alignment & Holistic integration of the audio and visual streams: how naturally the music complements the video across temporal, thematic, emotional and structural correspondence. \\
\bottomrule
\end{tabular}
\caption{The seven axes along which Gemini scores audio--visual alignment as defined in Video-Robin \cite{lokegaonkar2026videorobinautoregressivediffusionplanning}. Each axis is rated independently on the same scale; \textit{Overall Alignment} is a separate holistic judgement rather than an aggregate of the preceding six.}
\label{tab:gemini_axes}
\end{table*}

\end{document}